\documentclass[11pt,twoside]{article}
\usepackage{verbatim,epsfig}
\leftline{\copyright~ 1999 International Press}
\leftline{Adv. Theor. Math. Phys. {\bf 2} (1999) {\ 1183 -- 1247} }

\def\ut#1{\rlap{\lower1ex\hbox{$\sim$}}{#1}}

\def\SU{{\rm SU}}

\def\SO{{\rm SO}}

\def\R{{\rm I}\!{\rm R}}
\def\M{{\cal M}}
\def\A{\underline{A}}

\def\w{w}

\newcommand{\pic}[5]{\raisebox{#3pt}{
\hspace{#4pt}\psfig{file=#1.eps,height=#2pt,silent=}\hspace{#5pt}}}

\newcommand{\kd}[1]{\mathchoice{%
\pic{#1}{60}{-30}{2}{2}}{
\pic{#1}{13}{-6}{-3}{2}}{ 
\pic{#1}{9}{-2}{-3}{1}}{
\pic{#1}{7}{-1}{-1}{0}}}

\newcommand{\KD}[1]{\mathchoice{%
\pic{#1}{120}{-60}{2}{2}}{
\pic{#1}{50}{-25}{-3}{2}}{ 
\pic{#1}{9}{-2}{-3}{1}}{
\pic{#1}{7}{-1}{-1}{0}}}

\begin{document}
\begin{center}
\vspace{0.2in}
{\huge \bf Spin Foam Models and \\[5mm] the Classical Action Principle}
\vspace{0.2in}

{\bf Laurent Freidel$^{a}$
, Kirill Krasnov$^{b}$}
\vspace{0.2in}

$^{ab}$Center for Gravitational Physics and Geometry \\
Department of Physics, \\
The Pennsylvania State University \\
University Park, \\
PA 16802, USA\\
{\tt freidel@phys.psu.edu}\\
{\tt krasnov@phys.psu.edu}
\vspace{0.1in}

$^{a}$ Laboratoire de Physique Th\'eorique ENSLAPP \\
Ecole Normale Sup\'erieure de Lyon \\
46, all\'ee d'Italie,\\
69364 Lyon Cedex 07, France\\
\renewcommand{\thefootnote}{}
\footnotetext{\small e-print archive: {\texttt http://xxx.lanl.gov/abs/hep-th/9807092}}
\renewcommand{\thefootnote}{\arabic{footnote}}
\vspace{0.2in}
\end{center}
\begin{abstract}
We propose a new systematic approach that allows one to derive
the spin foam (state sum) model of a theory starting from the
corresponding classical action functional. It can be
applied to any theory whose action can be written as that of
the BF theory plus a functional of the B field. 
Examples of such theories include
BF theories with or without cosmological term, Yang-Mills theories 
and  gravity in various spacetime dimensions.  
Our main idea is two-fold.  First, we propose to take into account in 
the path integral certain distributional configurations of the B 
field in which it is concentrated along lower dimensional hypersurfaces
in spacetime.  Second, using the notion of generating functional  we develop 
perturbation expansion techniques,
with the role of the free theory played by the BF theory.  
We test our approach on various theories for which the corresponding spin foam 
(state sum) models are known. We find that it exactly reproduces the known
models for BF and  2D Yang-Mills theories.
For the BF theory with cosmological term in 3 and 4 dimensions
we calculate the terms of the
transition amplitude that are of the first order 
in the cosmological constant, and find an agreement with
the corresponding first order terms of the known state sum models. We 
discuss implications of our results for existing quantum gravity models.
\end{abstract}
\newpage
\headheight 0.5in
\textheight 8in
\pagenumbering{arabic}
\setcounter{page}{1184}

\pagestyle{myheadings}
\markboth{\it SPIN FOAM MODELS ...}{\it L. FREIDEL, K. KRASNOV}

\section{Introduction}

Recently Baez \cite{SpinFoams}, motivated by results \cite{RR} of the
canonical (loop) approach to quantum gravity, introduced a notion
of ``spin foam models''. The spin foam model of a theory, 
as we explain in details below, gives a way to calculate 
transition amplitudes for this theory. A transition amplitude
arises as a sum over branched colored surfaces, or, using
the name proposed by Baez, spin foams.  The results of Reisenberger 
and Rovelli \cite{RR}, Reisenberger \cite{Mike97}, Barrett and Crane 
\cite{BC} indicate that quantum gravity might be constructed as a spin 
foam model.

However, spin foam models arise not only in the
context of canonical (loop) quantum gravity. Indeed, many of
the known TQFT's --- for example, the Migdal-Witten model in two 
dimensions \cite{Migdal,Witten}, the Ponzano-Regge \cite{PR} and
Turaev-Viro \cite{TV} models in three
spacetime dimensions, the Ooguri model \cite{Ooguri} and its 
generalizations by Crane, Kauffmann and Yetter \cite{CKY} --- 
are state sum models and, by considering instead of
sums over colored triangulations sums over the 
corresponding colored dual complexes, can be formulated
as spin foam models.
The state sum models mentioned are believed to correspond to 
certain well-known topological field theories. However,
a precise relation between the classical theory and the corresponding
state sum model is not known for some of these models. Moreover,
even for the models whose relation to the corresponding classical
theory is established, this relation is 
rather indirect. Thus, summarizing, one can say that there
exists a large class of spin foam models that are claimed
to correspond to such classical theories as Yang-Mills, BF theories
and gravity in various dimensions, but no precise relation to
this effect is known in many cases.

The present paper is devoted to the study of this relation. More precisely,
we study the relation between a spin foam
formulation of a theory and the corresponding classical action
principle. We propose and study a particular procedure that allows
one to get a spin foam formulation for a large class of
theories starting from the corresponding action principle.
Our approach is not limited to any particular
spacetime dimension. However, having in mind
applications to the models mentioned above, we restrict ourselves to
two, three and four spacetime dimensions.\footnote{For an application of our procedure to higher-dimensional gravity see
\cite{FKP}.}

When considering gravity, for simplicity, we
restrict ourselves to metrics of Euclidean signature. 

Our approach is inspired first of all by the results
of the canonical (loop) approach \cite{RS,SpinNet,Measures,QGeom}
to quantum gravity. Of great importance for us were also 
the ideas of Regge \cite{Regge}, the work of Baez 
\cite{BaezDegenerate} on degenerate solutions of general relativity
and some results of Barrett \cite{Barrett}.
 Motivated by all these results we propose to ``approximate'' some of the 
 continuous fields appearing in the Lagrangian of the theory by 
 distributional fields concentrated along lower 
 dimensional hypersurfaces in spacetime. 
To make contact with the known lattice models, we put our 
 distributional fields on a triangulated manifold, so that they are 
 concentrated along polygons of the dual complex.  However, the idea is 
 also valid in a context more general than a lattice model.  The action 
 of the theory calculated on such distributional fields becomes a sum 
 of certain integrals over the hypersurfaces of lower dimension.  
 Then, to calculate a transition amplitude for the theory, one 
 constructs the path integral of the exponentiated action over such 
 distributional fields.  As we shall see, this path integral becomes 
 quite tractable.

To make this idea little more precise, let us specify the
class of theories that can be treated within our approach.
We consider theories whose action is BF-like, that is, 
whose action is of the form
\[
\int_\M \left[ {\rm Tr}\left( E\wedge F \right) + \Phi(E)\right],
\]
where $E$ is the Lie algebra valued two-form field 
(for instance, $B$ field of BF theory), 
$F$ is the curvature of the connection form and $\Phi$ is
certain (polynomial) function of the $E$ field, which can
also depend on some Lagrange multipliers, as in the 
case of gravity in four dimensions, or on an additional
background structure, as, for example, a fixed measure on 
$\cal M$ in the case of Yang-Mills theory in two
dimensions. Below we will give
many examples of theories belonging to this class. Thus, the
action is that of the BF theory with an additional term. We call
the B field E because of its relation with the non-abelian 
``electric'' field of the canonical formulation.

In order to compute a transition amplitude for a theory belonging to 
this class, we take the path integral over classical configurations where 
the $E$ field, which is a zero-, one- or two-form in different dimensions, is  
a distributional field that vanishes everywhere except on two dimensional 
polygons in spacetime.  Note that in our approach only these distributional
configurations of the E field are taken into account in the path integral. No 
usual smooth configurations of the E field are summed over.
This can be viewed as a discretization procedure where  we 
``approximate'' the smooth $E$ field by ``squeezing'' it into a 
distributional field on a collection of two-dimensional polygons.  One 
can expect that this approximation becomes good as the two-dimensional 
polygons become sufficiently packed in spacetime.
However, this procedure can also be justified at  a 
deeper level, as we shall argue.

First, a justification for considering distributional fields
in the path integral comes from the canonical quantum
theory. The phase space of all theories we work with,
as it can be seen from the above action functional, is that
of Yang-Mills theory. The space of quantum states
of the canonical theory is then given by $L^2({\cal A}/{\cal G})$,
where ${\cal A}/{\cal G}$ is the space of connections on the 
spatial manifold modulo gauge transformations, and $L^2$
is defined by an appropriate choice of measure on this space.
In the case the spatial manifold is one-dimensional, there
exists quite a natural choice of this measure: the functionals
on ${\cal A}/{\cal G}$ are just class functions, and it is
natural to use the Haar measure on the group to define the
inner product of such functionals. In higher dimensional cases, 
an analogous construction of $L^2({\cal A}/{\cal G})$ was
developed \cite{Measures} within the approach of 
canonical (loop) quantum gravity. In this construction, 
integrable functions arise by considering the so-called cylindrical
functionals, that is, functionals that depend
on connection only through holonomies along some paths in space.
In other words, one constructs the space of states of the theory
by considering spaces of states of lattice gauge theories on
graphs in space, and then taking the projective limit over
graph refinements (see \cite{Measures} for details). A characteristic
feature of the quantum theory constructed with such a space of
states is the distributional nature of the electric field
operator. Let us illustrate this on the case of (2+1)
dimensional theory. A typical state from $L^2({\cal A}/{\cal G})$
depends on the connection only through holonomies along some paths
in space. Although this is not important for our 
heuristic consideration here, let us note that gauge 
invariance implies that the paths form a closed
graph in space. Let us denote the edges of such a graph by $e_i$.
Then a typical state supported on this graph is of the
form
$$\Psi\left(h_{e_1}(A),\ldots,h_{e_n}(A)\right),$$
where $h_{e_i}(A)$ is the holonomy of $A$ along the 
edge $e_i$, and $\Psi$ is a function with sufficiently good 
behavior. The electric field $E$, which classically
is a canonically conjugate quantity to $A$, quantum 
mechanically becomes the operator $\delta/\delta A(x)$
of variational derivative with respect to $A$. This operator,
when acting on a typical state, gives zero for all points $x$
except the points lying on the edges $e_i$, where the
result is distributional. Thus, in a typical quantum state,
the electric field $E$ is distributional with support on
edges $e_i$ of the corresponding graph. 
\begin{figure}

\centerline{\psfig{file=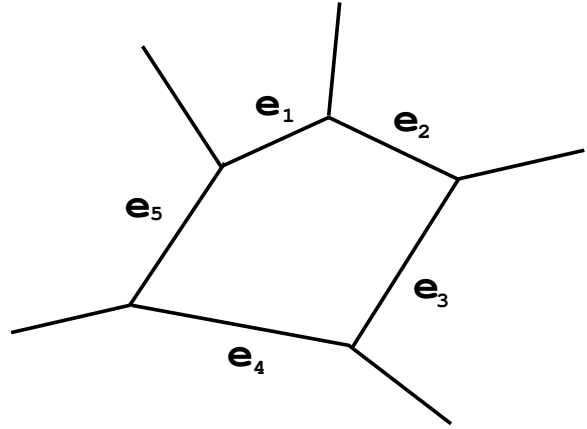,width=3.0in}}

\caption{A typical quantum state of (2+1) dimensional
theory is labelled by a graph in space. The corresponding
functional of connection depends on $A$ only through holonomies
along edges $e_i$. The electric field in such state is 
distributional and concentrated along the edges $e_i$.}
\end{figure}
One normally views a state of the canonical theory as a state
``at a given time''. Thus, electric field ``at each given
time'' is distributional and concentrated along edges of the
graph. This suggests that one should think of the ``history''
of the electric field configuration as of a collection of 
{\it two-dimensional} surfaces in spacetime, 
along which the electric field is distributional. When 
sliced by a spatial hypersurface ``at given time'', such a 
collection of surfaces gives a graph in space, and
electric field is distributional along the edges of this
graph. These heuristic considerations suggest that
one should consider configurations of $E$ field that
are distributional and concentrated along 
two dimensional surfaces in spacetime.  As we will see
in more details below, the surfaces of support of $E$
field are always two-dimensional, independently of the
dimension of the spacetime manifold.

The second justification for considering  the path 
integral over such distributional $E$ fields comes from the fact of 
existence of similar distributional solutions of the equations of 
motion in the classical theory.  As it was shown by Baez 
\cite{BaezDegenerate} for the case of gravity, there exist solutions 
of the classical equations of motion in which the $E$ field is 
distributional and concentrated along two-dimensional surfaces in 
spacetime.  In these solutions the $E$ field is zero except in a 
tubular neighborhood of a two-dimensional surface in spacetime.  The 
connection field away from the surface then must be either arbitrary 
or flat, depending on the detailed form of the Lagrangian of the 
theory.  Solutions of this type can be shown to exist for a large 
class of theories.  Here we take the fact of the existence of such 
solutions as a motivation for our construction.  Namely, we know that 
the simplest way to approach the computation of a path integral is to 
consider the semi-classical approximation.  Then the Feynman 
path integral is evaluated as an integral over the classical moduli 
space (space of classical solution) with a measure given by the one 
loop expansion of the action around the classical solution. In this
case, there may be different phases in the theory, and 
the type of solutions that dominate the path integral may be 
different in different phases. Thus, there may 
exist a phase in a theory of the type we study where the path
integral is dominated by the distributional solutions found by
Baez. At this point it is instructive to make an analogy with the
usual Yang-Mills theories. In (abelian) Yang-Mills theories
one also finds ``distributional'' solutions of the classical equations of 
motion: monopole solutions.  In Dirac's way \cite{Dirac}
of treating the monopole, it is described by a 
configuration of fields in which magnetic field is distributional 
along certain lines.  Distributional monopole 
solutions are not in the spectrum of perturbative quantum theory.  
However, one must include them when calculating the partition function 
of the theory via path integral method to reproduce correctly 
properties of certain phases of the theory, such as the confining 
phase.  In fact, the confining phase of the theory is the one in which 
monopole solutions dominate the path integral \cite{Pol}.  
Distributional solutions of the type discovered by Baez 
\cite{BaezDegenerate} are very similar in nature to monopole 
solutions.  In fact, they are close to dual monopoles: electric field 
is distributional and concentrated along flux lines in space, instead 
of the magnetic field.  Thus, the analogy with monopole solutions 
suggests that one should include these distributional solutions when 
taking the path integral.  In this paper we include {\it only} these 
distributional solutions, neglecting the usual smooth solutions 
playing the role in perturbative quantum theory.  Thus, we attempt to 
study the phase of various theories in which the distributional 
solutions dominate the path integral.  This is the motivation for 
considering the path integral over distributional configurations of 
$E$ field.

The last but not least justification of our approach is the 
test of the procedure on the exact results 
coming from TQFT's.  As we shall see, for a large class of theories
our approach gives results that are in a good agreement with the known 
state sum models.

In this paper we only calculate path integrals over the 
distributional fields ``living'' on a fixed collection of surfaces in 
spacetime.  In the case of topological theories, the result for a 
particular fixed collection of surfaces does not depend on the 
collection.  One cannot expect this property to hold for such theories 
as gravity, which is not a topological theory.  Thus, one might want 
to perform a sum over all possible collections of surfaces in 
spacetime, or take a limit as the triangulation becomes more refined.  
We do not address this, very important, problem in the present paper.

Usual interactive quantum field theories have been successfully 
understood using perturbative expansion in terms of Feynman graphs.
In this case the interacting QFT is considered as a perturbation 
of a free field theory.  From a technical point of view, what makes this 
approach successful is the exact solvability of the free field theory, i.e., 
our ability to compute all possible correlation functions.  This 
ability can be recasted into the knowledge of the generating 
functional, and Feynman graphs can be viewed as the evaluation of certain 
differential operators acting on the generating functional.

We will employ a somewhat similar strategy.
In order to calculate transition amplitudes for our class of theories we 
will use the machinery of generating functionals.  In our approach the role
of the ``free'' field theory is played by the BF theory and the terms in the 
action polynomial in $E$ are analogous to the ``interaction'' terms.
The idea to use the BF theory as a ``free'' field theory is not new: it has
been applied with some success in the 
context of Yang-mills theories (see \cite{Martellini}). 
However, the details of how the BF theory is used in our approach 
differ from those of \cite{Martellini}. In 
order to calculate a transition amplitude, i.e., the path integral 
of the exponentiated action
\[
\int {\cal D}A {\cal D}E \,\,e^{i\int [{\rm Tr}\left( E\wedge F\right) + \Phi(E)]},
\]
 we formally rewrite this path integral as
\[
\left(e^{i\int \Phi\left({\delta\over i\delta J}\right)}\,Z[J]\right)_{J=0},
\]
where
\[
Z[J]:=\int {\cal D}A {\cal D}E \,\,e^{i\int [{\rm Tr}\left( E\wedge F\right) + 
{\rm Tr}\left( E\wedge J\right) ]}.
\]
We shall refer to $Z[J]$ as the {\it generating functional}. Here $J$
is a Lie algebra valued two form field.
Using terminology from field theory we will call the $J$ field 
{\it current}.
One of our main results is the exact computation, 
in the  context of spin foam models,
of the  generating functional $Z$.
It is obtained by integrating over fields $A,E$ that ``live'' on a fixed 
triangulation of the spacetime manifold.  Then the transition amplitude 
for any theory of the type we consider is given as a formal power 
series in variational derivatives $\delta/\delta J$ with respect to 
the current field. The series we get are quite reminiscent of the usual 
Feynman diagram expansions.
Thus, the powerful technique of generating functionals 
allows us to study different theories --- such 
as BF theory and gravity --- from the same point of view.  We note 
that, to the best of our knowledge, the technique of generating 
functionals has never before been used in the context of spin foam 
models.

In cases when the spacetime manifold is two, three and four-dimensional,
for each fixed triangulation we find an explicit expression 
for the generating functional. We then use the obtained expressions
to calculate the 
transition amplitudes for various theories.  We find that our approach 
exactly reproduces the state sum models of 2D Yang-Mills and  BF theory in any 
dimension.  For BF theories with cosmological term we compare the 
amplitude obtained within our approach with the one given by the 
corresponding state sum model by decomposing them into the power 
series in the cosmological constant and comparing the first order 
terms.  Again, we find an agreement.  We make some comments as to 
consequences of our results for the quantum gravity models in four 
spacetime dimensions.

The paper consists of two main parts. In the first part, containing this
section and  Sec.\ref{sec:gf}, we
introduce the path integral techniques in the context
of spin foam models. Here we also introduce  the generating 
functional, and compute it in various spacetime dimensions.
The second part contains applications of the general
techniques developed earlier to concrete theories.  
This also serves as a test of our approach, for one can
compare the spin foam models obtained by our techniques
with the known state sum models. In Sec.\ref{sec:th} we
define the class of theories that can be treated by 
techniques described in this paper, and give examples
of physically interesting theories belonging to this class.
Sec.\ref{sec:sf} describes briefly some known state
sum models and summarizes the known arguments that relate
the state sum models to the classical theories. Here
we also obtain some properties of various state sum models
that will be needed later. Finally,
using the generating functional technique, we derive in 
Sec.\ref{sec:appl} the 
spin foam models corresponding to various theories of Sec.\ref{sec:th}
and compare them with the known
models. We conclude with a discussion.

\section{Generating functional}
\label{sec:gf}

As we briefly explained in the Introduction, to calculate transition amplitudes
for a theory of the type we consider here it is very convenient to first calculate the
generating functional of BF theory. Then transition amplitudes can be obtained 
as formal power series in derivatives with respect to the current. 
In this section we calculate the generating functional of the BF theory,
i.e., the path integral
\begin{equation}\label{genf}
Z[J]=\int {\cal D}A {\cal D}E \,\,e^{i\int [{\rm Tr}\left( E\wedge F\right) + 
{\rm Tr}\left( E\wedge J\right) ]}
\end{equation}
in two, three and four spacetime dimensions. 
Here $A$ is a connection on some $G$ bundle over the spacetime manifold 
$\cal M$, where $G$ is the 
gauge group of the theory, $F$ is the curvature 2-form of the connection, and 
$E, J$ are Lie algebra valued (d-2)-form fields, where d is 
the dimension of $\cal M$.
For simplicity,
we consider here only the case when the spacetime manifold $\cal M$
is closed.  We also restrict ourself to the case of the gauge group being 
$G=\SU(2)$. 
Although for practical applications of spin foam models one is usually 
interested in the case when $\cal M$ has a boundary consisting of 
``initial'' and ``final'' spatial hypersurfaces, we do not consider 
this here because our main goal is to compare the spin foam models 
obtained by our techniques with the known state sum models, which can 
be done already in the case of closed $\cal M$.  At any rate, the 
inclusion of boundaries is straightforward, and all our formulas can 
be easily generalized to this case.  To give meaning to the path 
integration over spacetime fields $A,E$, we replace the $E$ field by 
certain distributional field concentrated over two-dimensional 
surfaces in spacetime.  To make contact with the known state sum 
models we put our fields on the 2-dimensional cellular complex dual to 
some triangulation of the spacetime manifold.  In each particular 
calculation this triangulation is fixed and all results explicitly 
depend on it.  We first investigate the generating functional in a 
general case, without specifying the dimension of spacetime manifold, 
and then find an explicit expression for it in each particular 
dimension.

\subsection{General framework}

Let us fix a triangulation $\Delta$ of a d-dimensional compact 
oriented spacetime manifold $\M$.  This triangulation defines another 
decomposition of $\M$ into cells called dual complex.  There is 
one-to-one correspondence between $k$-simplices of the triangulation 
and (d--k)-cells in the dual complex.  We orient each cell of the dual 
complex in an arbitrary fashion, which also defines an orientation for 
all simplices of the triangulation.  The (d--2)-form $E$ can now be 
integrated over the (d--2)-simplices of the triangulation the result 
being a collection of the Lie algebra elements $X$ --- one vector 
$X\in{\rm su}(2)$ for each 2-cell dual to a (d--2)-simplex of the 
triangulation.  We would like to discretize our model by replacing the 
continuous $E$ field by the collection of the Lie algebra elements 
$X$.  It turns out, however, that this is not yet the most convenient 
set of variables for the theory.  For reasons which we give below, we 
use another, more convenient set of variables introduced by 
Reisenberger \cite{Mike96}.

 In what follows we will often call 2-cells of the dual 
triangulation {\it dual faces}.
 Instead of a single $X$ for each dual face,
we introduce a set of variables, which we 
call $X_w$. To do this, we divide each dual face into
what Reisenberger calls ``wedges'' (see Fig. 2). To construct wedges of the
dual face one first has to find the ``center'' of the dual face. 
This is the point on the dual face where it intersects with
the corresponding (d--2)-simplex of the triangulation.
One then has to draw lines connecting this center with the centers of the 
 neighboring  dual faces.  The part of the dual face that lies between 
 two such lines is exactly the wedge. Thus, each dual face 
 splits into wedges, and we assign a Lie algebra element $X_w$ to each 
 wedge $w$.  All variables $X_w$ that correspond to a single dual face 
 are required to have the same length. Recall that vectors of the same length
 only differ by a gauge transformation.  Wedges of a given dual face are 
 in one-to-one correspondence with the d-simplices of the triangulation 
 neighboring this dual face.  Thus, the physical meaning of each 
 variable $X_w$ can be said to be the integral of $E$ over the (d--2) 
 simplex of the triangulation ``from the point of view'' of a 
 particular d-simplex containing this (d--2)-simplex.  In other words, 
 with the introduction of the wedge variables $X_w$ we allow integrals 
 of $E$ over one and the same (d--2)-simplex of triangulation to be 
 different when this simplex is considered as belonging to different 
 d-simplices, as long as lengths of all such $X_w$ coincide.  Note that 
 the number of variables $X_w$ that arises this way for a given 
 triangulation is equal to the number of d-simplices times the number 
 of (d--2)-simplices in each d-simplex.
\begin{figure}
\centerline{\epsfig{figure=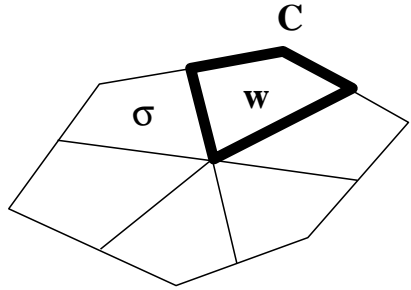,height=1.5in}}
\bigskip
\caption{A face $\sigma$ of the dual triangulation. The portion of 
$\sigma$ indicated by bold lines is what is called wedge here. The point
labelled by $C$ is the center of one of the d-simplices 
neighbored by $\sigma$.}
\end{figure}

Having discussed the geometrical meaning of the wedge variables
$X_w$ we are ready to introduce the distributional $E$ fields. 
Heuristically, our procedure of replacement of a smooth 
$E$ field by a distribution concentrated along the wedges
amounts to ``squeezing'' of the smooth $E$ field that
is ``spread'' over a (d-2) simplex of the triangulation to a
single point on this (d-2) simplex, the point where the
simplex intersects with the 2-cell of the dual complex.
Thus, we define a distributional field $E_w$
concentrated along a wedge $w$
to be a 2-form satisfying the following relation:
\begin{equation}\label{distr}
\int_\M {\rm Tr}(E_w \wedge J) = {\rm Tr}(X_w \int_w J).
\end{equation}
Here $J$ is any ${\rm Ad}(P)$-valued 2-form, and $w$
stands for a wedge. The integral on the
right-hand-side is performed over the wedge $w$.
This, in particular, implies that the Lie algebra element $X_w$ 
is equal to the
integral of $E_w$ over the (d--2)-simplex of the triangulation
that is dual to the 2-cell $\sigma$. We then define distributional
field $E$ to be 
\begin{equation}\label{distr-1}
E = \sum_w E_w.
\end{equation}

Although with the introduction of the wedge variables
we have increased the number of independent variables in the theory,
one can still argue that the physical content of the theory should
be the same because the norm of all the $X_w$ belonging to one and the
same dual face are required to be equal. For the case of BF theory  
one can directly prove that the introduction of the 
independent wedge variables does not change the theory
in the case the topology of dual faces $\sigma$ is that of
a disc, which is what we will assume here.  
For more complicated theories with non-zero ``interaction''
term (the term in the action that makes the theory to be 
different from the BF theory), the wedge variables are so
indispensable that it is not clear whether one can 
even define the theory without these variables.  
It is interesting, however, that, even in 
 the case of BF theory one {\it has} to use the wedge variables if 
one wants to reproduce a triangulation-independent state sum model for 
the case the topology of $\sigma$ is different from that of a disc.  
This serves as a strong motivation for considering the wedge 
variables.  Another motivation for using them is, as we explain 
below, that the wedge variables appear naturally when one tries to 
implement the gauge symmetry in the theory.    Also, as we shall 
see, using these variables we will be able to reproduce the well-known 
state sum models in various dimensions.  Thus, we are going to use the 
set of wedge variables motivated, in particular, by the results we 
obtain.

The discretization we consider is consistent with the results of the
canonical approach. The restriction of $E$ constructed as described above 
to a generic ``spatial'' ((d--1)-dimensional) slice $\Sigma$ of $\M$ is 
non-zero only along edges of the graph $\Gamma$ in $\Sigma$, where 
$\Gamma$ is the intersection of the dual complex with $\Sigma$.  Thus, 
this ``discretization'' realizes rather explicitly Faraday's idea of 
lines of electric force.  Indeed, the restriction of $E$ on a spatial 
hypersurface can be described as a collection of flux lines of 
non-abelian electric field.  It is not hard to see what states would 
correspond to such $E$ in the quantum theory.  Recall that the quantum 
states are functionals $\Psi[\A]$, where $\A$ is the pullback of the 
connection field $A$ on the spatial hypersurface $\Sigma$.  The 
densitized vector field $\tilde{E}$ (that is dual to $\underline{E}$) 
becomes the operator of variational derivative with respect to $\A$.  
In order for $\tilde{E}$ in a state $\Psi[\A]$ to be zero everywhere 
except along the edges of $\Gamma$ the state $\Psi$ should depend on 
the values of the connection $\A$ only along the edges.  Note also 
that our construction of the distributional $E$ field gives us 
$\tilde{E}$ whose length is constant along each edge.  This means that 
the state $\Psi$ can depend on the values of $\A$ only as a function 
of the holonomies of the connection along the edges.  These are 
exactly the states one finds in the canonical approach to quantum 
theory, as we explained in the introduction.

To calculate the generating functional $Z$ as a function of a fixed 
triangulation $\Delta$, we have to take the integral of the exponentiated action over 
the ``discretized'' dynamical fields $A,E$, that is, the fields 
``living'' on the triangulation $\Delta$.  The ``discretized'' 
action, that is, the action evaluated on the distributional field $E$, 
becomes a function of the Lie algebra elements $X$, and a functional 
of connection $A$ and current $J$.  Using 
(\ref{distr}),(\ref{distr-1}), we find that this discretized action is 
given by
\begin{equation} \label{dis}
\sum_{w} \int_{w} \left[{\rm Tr}(F\,X_w)+{\rm Tr}(J\,X_w)\right],
\end{equation}
where the sum is taken over the wedges of faces of the dual complex
(`w' stands here for a wedge), 
and the integral is performed over each wedge, the
integrand being the curvature of the connection $A$ contracted with the
Lie algebra element $X_w$ ``living'' on that wedge plus the current $J$
contracted with $X_w$. Each 
integral is performed using the orientation of the dual face to which
the wedge belongs.

We could now substitute this discretized action into the path integral and
integrate over $X$ and $A$. However, we first have to discuss
what measure has to be used to integrate over $X,A$. 
To give meaning to the integration over $A$, let us
replace a continuous field $A$ by 
a collection of group elements. To do this we use the following 
approximation
\begin{equation} \label{approx}
\int_w {\rm Tr}(F\,X_w) \approx {\rm Tr}(Z_w X_w),
\end{equation}
where $Z_w$ is the Lie algebra element corresponding to the holonomy
of $A$ around the wedge (see Fig.  3). The base point of the 
holonomy is not fixed at this stage (see below).  In other words,
\begin{equation} \label{z}
\exp{Z_w} = g_1 h_1 h_2 g_2,
\end{equation}
\begin{figure}
\centerline{\epsfig{figure=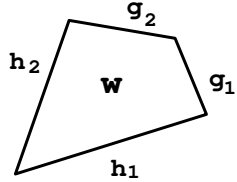,height=1in}}
\bigskip
\caption{A wedge $w$ of the dual face and the group elements:
holonomies along the edges of $w$.}
\end{figure}
where $g_1,h_1,h_2,g_2$ are the holonomies of $A$ along the four edges 
that form the boundary of the wedge $w$. We assume
a local trivialization of the bundle over $w$ so that the holonomies are 
group elements. The order in which the product of group elements is
taken is determined by the orientation of the dual face.
Of course, $\exp{Z_w}$ is defined only up to its conjugacy class
(any of the four edges can be taken to be the ``first''). Thus, 
there is an ambiguity in the choice of the base point for the
holonomy, which we have
to fix in some way. To fix it we define a notion 
of ``discretized'' gauge transformation and fix the
ambiguity requiring the discrete action to be gauge invariant.

But first, let us replace the continuous current field $J$
by a collection of Lie algebra elements $J_w$:
\begin{equation}\label{jw}
J_w:=\int_w J.
\end{equation}
Here, to perform the integration, a trivialization of the 
bundle over the wedge $w$ is chosen. The discretized
action now becomes
\begin{equation}\label{disw}
\sum_w {\rm Tr}(Z_w\,X_w)+{\rm Tr}(J_w\,X_w).
\end{equation}
We fix a definition of $Z_w$ in such a way that this
action is gauge invariant.

We define the discrete gauge transformation so that it
``acts'' in the center of each d-simplex. More precisely, 
a gauge transformation is parameterized by a collection
of group elements: a group element $g$ for each 
d-simplex. First of all, the gauge transformation is defined
for the holonomy $U$ of the connection along any loop that
starts and finishes at the center $C$ of the d-simplex. 
The transformation is as follows:
\begin{equation}\label{g-discr1}
U \to g U g^{-1}.
\end{equation}
The wedge variables $X_w$ and the discrete current
variables $J_w$ transform as 
\begin{equation}\label{g-discr2}
X_w \to g X_w g^{-1}, \quad J_w \to g J_w g^{-1}.
\end{equation}
With this definition ${\rm Tr}(X_w Z_w)$ is gauge invariant 
only when $\exp{Z}$ is defined as the holonomy around the wedge
$w$ whose starting and final point is the center $C$
of the d-simplex, as in (\ref{z}). This fixes the ambiguity in
$Z_w$. With $Z_w$ defined this way the discretized action
is gauge invariant.

The approximation (\ref{approx}) is good for $Z_w$ close to 
zero element in the Lie algebra. Thus, this approximation
is certainly justified for BF theory, where we expect 
only connections close to flat to matter
in the quantum theory. It is harder to justify this
approximation for theories, for which the classical
equations of motion do not imply the connection to
be flat, as is the case, for example, for BF theory
with cosmological term or for gravity. 
However, even for such theories, one would expect 
the approximation (\ref{approx}) become better as
the triangulation of the manifold becomes finer and the
dual faces become smaller. In quantum gravity we expect
finer triangulations to matter most, which can serve
as a justification for the above approximation in 
the case of gravity. The approximation (\ref{approx})
will be additionally justified when we discuss the
problem of the integration over $X_w$ variables.
To define the later we will refer to a certain standard
field theory calculation for BF theory in two dimensions.
As we shall see, the approximation (\ref{approx}) is
quite natural from the point of view of 2D BF theory.

The approximation (\ref{approx}) finishes the discretization procedure 
for the classical action.  We can now calculate the path integral for 
the generating functional $Z$ by integrating the exponentiated 
discrete action (\ref{disw}) over the Lie algebra elements $X_w$ and 
group elements $g,h$.  This path integral is given by
\begin{equation} \label{pathintBF}
Z(J,\Delta) = \prod \int_{\SU(2)} dg \prod \int_{\SU(2)} dh
\int \prod_{w}dX_w \, e^{i\,\sum_{w} 
{\rm Tr}(X_w Z_w)+{\rm Tr}(X_w J_w)}.
\end{equation} 
Here the integrals are taken over all group elements $g,h$,
entering the expression through $Z_w$, see (\ref{approx}).
These integrals form the discrete analog of ${\cal D}A$, $dg$ is
the normalized Haar measure $\int dg = 1$ on $\SU(2)$. 
The integrals over $X_w$ present here --
one for each edge -- form the analog of the integral over ${\cal D}E$. 
The measure $dX$ here is some measure on the Lie algebra.
For now, we will leave this measure unspecified. 
Recall now that the integrals over $X_w$
are not independent: all $X_w$ that belong to one and the same 
dual face should have the same norm. However, as we shall see,
there is no need to impose these constraints: one can integrate
over all $X_w$ independently, but later, when one integrates over
the group elements, only the contributions that come from $X_w$
of an equal norm will survive. Thus, we do not impose these
constraints in (\ref{pathintBF}). 

Let us now investigate the structure of the path integral 
(\ref{pathintBF}).  To calculate $Z$ we have to find the function of 
$\exp{Z_w},\exp{J_w}$ that is given by
\begin{equation} \label{1}
\int dX_w e^{i\,{\rm Tr}(X_wZ_w+X_wJ_w)}.
\end{equation}
In fact, it is not hard to see that this function is proportional to
the $\delta$-function of $\exp{Z_w}$ peaked at $\exp{J_w}$. The
proportionality coefficient can be a gauge invariant function of 
$J_w$.  As we explain below, this function must be set to be equal to 
$P(J_w)$, where $P$ is the function that relates Lebesgue measure on 
the Lie algebra and Haar measure on the group (see the Appendix B).  
Thus, as the result of (\ref{1}) we get:
\begin{equation} \label{2}
P(J_w)\,\delta(\exp{Z_w}\exp{J_w}),
\end{equation}
where $\delta$ is the standard $\delta$-function on the group.
Thus, the function of $Z_w,J_w$ given by (\ref{1}) can be written as
\begin{equation}
P(J_w) \sum_{j} {\rm dim}_{j}  \chi_{j}(\exp{Z_w}\exp{J_w}),
\end{equation}
where we used the 
well-known decomposition of $\delta$-function on the
group into the sum over characters 
$\chi_j(\exp{Z_w}\exp{J_w})$ of irreducible representation.
Here the sum is taken over all irreducible representations
of $\SU(2)$ (labelled by spins $j$), and ${\rm dim}_j=(2j+1)$.

Thus, we find that
\begin{equation} \label{k2}
Z(J,\Delta) = \prod \int_{\SU(2)} dg \prod \int_{\SU(2)} dh
\prod_{w}  P(J_w)  \sum_{j_w} {\rm dim}_{j_w}
\chi_{j_w}(\exp{Z_w}\exp{J_w}).
\end{equation}
The integration over the group elements can now be easily performed
using the well-known formulas for the integrals of the products of 
matrix elements. However, let us first give a systematic
derivation of the result (\ref{2}). 

To find the result of the integral (\ref{1})
we relate it to a more complicated integral,
for which a precise result is known from 
2D BF theory calculation. Recall that so far the
measure $dX_w$ in (\ref{1}) is unspecified. The
following relation to 2D BF theory will also
serve the purpose of specifying this measure.
Let us restrict
our attention to a particular wedge $w$. We can 
restrict the bundle $\cal P$ to $w$ and get an
$\SU(2)$ bundle ${\cal P}_w$ over $w$. Let $A$ be
a connection on ${\cal P}_w$, and $E$ be an 
${\rm Ad}(\cal P)$ valued 0-form. Consider
the following path integral:
\begin{equation}\label{bf1}
 \int {\cal 
D}A {\cal D}E \exp\left( i\int_{w}\left[{\rm Tr}(F\,E)+ {\rm 
Tr}(J\,E)\right]\right),
\end{equation}
where the integration over $A$ is performed subject to
the condition that the connection on the boundary of $w$
is fixed, and $J$ is given by 
\begin{equation}\label{source}
J = \delta(p) J_w,
\end{equation}
where $p$ is an arbitrary fixed point on $w$, and
$J_w$ is the same as in (\ref{1}),(\ref{2}).
The path integral (\ref{bf1}) is just a partition
function of BF theory on the disk with the distributional source
given by (\ref{source}). This partition function can be
derived using results of \cite{Blau}.  The result is given by (\ref{2}), 
where $Z_w$ is the Lie algebra element that corresponds to the 
holonomy of $A$ along the boundary of $w$.  We will not present this 
calculation here.  Instead, we refer to the calculation performed in 
\cite{Blau} for the partition function of 2D BF theory on a punctured 
sphere.  The result (\ref{2}) can then be checked by taking the 
partition function on the disk (equal to $\delta(g)$, where $g$ is the 
holonomy along the boundary of the disk), and integrating it with 
(\ref{2}) over $dg$.  This must give the partition function on a 
punctured sphere, and indeed reproduces the result given in 
\cite{Blau}.  The only cautionary remark we have to make is that the 
calculation performed in \cite{Blau} finds a gauge invariant partition 
function, that is, the one in which one takes $J = \delta(p) hJ_w 
h^{-1}$ and integrates over $dh$.  The techniques developed in 
\cite{Blau} can be used only to calculate gauge invariant quantities, 
and are not directly applicable to the integral (\ref{bf1}).  Thus, 
strictly speaking, using the results of \cite{Blau} one can only argue 
that (\ref{bf1}) is equal to
\begin{equation}
P(J_w) \delta(e^{Z_w} h e^{J_w} h^{-1}),
\end{equation}
where $h$ is some group element. To get rid of
$h$ in this expression, and, thus, to get (\ref{2}),
we will recall how ``discretized'' gauge transformations
act on the Lie algebra elements $Z_w,J_w$, see
(\ref{g-discr1}),(\ref{g-discr2}). The result of 
(\ref{1}) must be invariant under this gauge
transformations. It is not hard to see that this
fixes $h$ above to be unity, thus, giving (\ref{2}).

Having discussed how one can calculate the path integral
(\ref{bf1}), let us now show that this path integral
is, in fact, equivalent to (\ref{1}). Indeed, the
integration over $E$ in (\ref{bf1}) can be performed
in two steps. First, one integrated over $E(x), x\not=p$,
then one integrates over $E(p)$:
\begin{equation}\label{bf2}
\int dX_w \int_{E(p)=X_w} {\cal D}E \int {\cal D}A
\exp\left( i\int_{w}\left[{\rm Tr}(F\,E)+
{\rm Tr}(J\,E)\right]\right).
\end{equation}
Since the current is distributional and concentrated at
point $p$, the last term in the exponential does not
matter when one integrates over ${\cal D}E, {\cal D}A$.
On the other hand, using the same approximation as the one used in
(\ref{approx}), it is not hard to show that 
\begin{equation}
\int_{E(p)=X_w} {\cal D}E \int {\cal D}A
\exp\left( i\int_{w}{\rm Tr}(F\,E)\right),
\end{equation}
where the integral over $A$ is taken with $A$
on the boundary of $w$ fixed, is approximated by
\begin{equation}
e^{i {\rm Tr}(Z_w\,X_w)},
\end{equation}
where $\exp{Z_w}$ is the holonomy of $A$ along the boundary of $w$.
Putting this back to (\ref{bf2}) one gets exactly (\ref{1}).
This finishes our discussion of the derivation of (\ref{2}).

Having discussed the derivation of the expression 
(\ref{k2}) for the generating functional, we can now perform
the integrals over the group elements. Integration over the group 
elements $h$ that correspond to edges dividing dual faces into wedges 
is the same in any dimension, and we perform it here.
The rest of the group elements corresponds to edges that form the 
boundary of the dual faces. The integration over these 
group elements $g$ is different in different dimensions,
and we perform it in the following subsections.
Each group element $h$ is ``shared'' by two
wedges; thus, we have to take the integral of the product of
two matrix elements. Such an integral is given by (\ref{a1}) of the
Appendix. Integrating over all these edges, and making a
simple change of variables to eliminate trivial integrations, 
we find
\begin{eqnarray} \label{k1}
Z(J,\Delta) = 
\prod_{\epsilon} \int_{\SU(2)} dg_\epsilon \times \\
\prod_{\sigma} \sum_{j_\sigma} ({\rm dim}_{j_\sigma})^{\kappa_\sigma} 
P(J_1)\cdots P(J_n) 
\chi_{j_\sigma}(\exp{J_1}\,g_{\epsilon_1}\,\exp{J_2}\cdots\exp{J_n}\,g_{\epsilon_n}). 
 \nonumber
\end{eqnarray}
Here the remaining integrals are over the group elements $g_\epsilon$ that
correspond to the edges $\epsilon$ of the dual complex (edges that connect 
centers of 4-simplices). The second product is taken over the dual
faces $\sigma$; $j_\sigma$ is the spin labelling the dual face $\sigma$, 
$\kappa_\sigma$ is the Euler characteristics of $\sigma$.  It is equal 
to unity in case the dual face has the topology of a disc.  In what 
follows we will always assume that this is the case.  The order of the 
group elements in the argument of $\chi_{j_\sigma}$ is clear from the 
Figure 4.
\begin{figure}
\centerline{\epsfig{figure=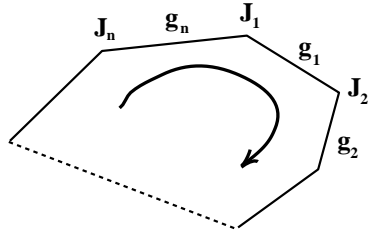,height=1.2in}}
\bigskip
\caption{The product of the group elements 
is the holonomy around the dual face $\sigma$ 
with the insertions of the group elements $\exp{J}$ at each center of the 
corresponding d-simplex.}
\end{figure}

Note that in the integration we just performed survived only the terms for
which the spins of all wedges corresponding to one and the same dual 
face are equal. In (\ref{k1}) this spin, which is the same for
all wedges belonging to the same dual face, is denoted by $j_\sigma$.
As one can see from the formula (\ref{IFtr}) for the
inverse Fourier transform, the spin labelling a wedge in
(\ref{k2}) has the meaning of the length of the corresponding
Lie algebra element $X$. Thus, as we said above, there is
no need to impose the constraint that length of all wedge
variables $X_w$ belonging to one and the same dual face
are equal: this constraint gets imposed automatically
when the integration over the group elements is
performed.

The expression (\ref{k1}) for the 
generating functional can be further simplified 
by integrating over the group elements $g_\epsilon$.
However, the result of this integration 
is different in different dimensions, 
and we will perform it separately for each particular case.

Before we calculate the generating functional for each particular
spacetime dimension, let us pause to discuss some general 
properties of the expression for $Z(J,\Delta)$.
The generating functional has several symmetries which we describe 
as follows. First, the generating functional is invariant 
under gauge transformations. Namely, let us consider a transformation $J_\w 
\rightarrow g_{h(w)} J_\w g_{h(w)}^{-1}$, where $h(w)$ denotes the 
d-simplex of $\Delta$ to which $w$ belongs. Here $g_{h(w)}$ is
the same for all $w$ belonging to the simplex $h$.
The generating functional 
(\ref{pathintBF}) satisfies:
\begin{equation}
Z(gJg^{-1},\Delta) = Z(J,\Delta).
\end{equation}
This ``discrete'' gauge transformation is parameterized by one
group element for each d-simplex of $\Delta$.
However the expression (\ref{k1}) for the generating functional has a bigger 
invariance.  Namely, let us denote by $e_+(w), e_-(w)$ 
the two edges 
of the wedge $w$ which meet at the center of the d-simplex $h(w)$. 
Let us associate with each edge $e_\pm(w)$ a group
element $g_{e_\pm(w)}$ so that it is one and the same for different
wedges $w$ when $e_\pm(w)$ is one and the same.
The transformation of all currents according to 
$J_w \rightarrow g_{e_+(w)}J_w g_{e_-(w)}^{-1}$ leaves the 
generating functional invariant. This transformation,
which is parameterized by $d+1$ group elements per simplex,
contains the gauge transformation as its particular case.
The later corresponds to $g_{e_+(w)}=g_{e_-(w)}=g_{h(w)}$
for all $w$ that belong to $h$. The appearance and significance of the 
described extra symmetry is not yet clear to us.

The generating functional is covariant under the diffeomorphism group.  
Let $f$ denote 
a diffeomorphism of the underlying manifold $\M$, and denote by $f(\Delta)$ 
the image of the embedded triangulation $\Delta$ under the diffeomorphism. 
If $J_\w =\int_\w J$, 
let us denote $f^*J\equiv \int_{f(\w)}J = \int_\w f^*J$.  Then
\begin{equation}
Z(J,f(\Delta)) = Z(f^*J,\Delta).
\end{equation}
The generating functional is generally not invariant under 
a refinement of the triangulation $\Delta$ unless the theory is a topological  
field theory (unless $J=0$).

To define the generating functional we had to choose an
arbitrary orientation of the wedges. However, the generating functional
is independent of the orientation chosen. If one, for example, reverses 
the orientation of the wedge $\w$  this results: 
(i) in the change $J_\w = \int_\w J$ into $-J_\w$; (ii) it changes
the holonomy of the connection along the wedge into its 
inverse.  The two effects cancel each other leaving  the generating
functional invariant.

\subsection{2 dimensions}
\label{sec:2d}

The case of two spacetime dimensions is somewhat special in the
sense that dual faces (2-cells) cover the manifold (see Fig. 5).
\begin{figure}
\centerline{\epsfig{figure=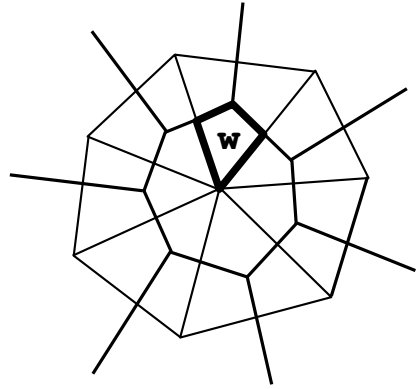,height=2in}}
\bigskip
\caption{Part of a triangulation of 2-dimensional spacetime
manifold, with one face of the dual complex shown. 
Bold lines indicate a wedge of the dual face.}
\end{figure}
The general construction of the previous subsection prescribes
to replace the continuous field $E$, which in the case of two
dimensions is a zero form, by a distributional field concentrated
along dual faces. However, since the spacetime is two-dimensional,
the dual faces cover the manifold, and we just have to replace the
$E$ field by a field constant on each wedge, and equal to the
Lie algebra element $X_w$. 

Let us now see that it is quite a reasonable thing to do from the
point of view of the canonical quantum theory. In the case 
$\cal M$ is compact, which is the case of interest for us here,
a spatial slice $\Sigma$ of $\cal M$ consists of a finite
number of circles $S^1$. The states of
the canonical theory are just class functions of the holonomies
of $\A$, which is the pullback of $A$ on $\Sigma$, around this circles.
The corresponding field $\underline{E}$, which in the quantum
theory becomes the operator of variational derivative with
respect to $\A$, has thus a constant norm along each disjoint component
of $\Sigma$. It thus makes sense to replace the $E$ field by
a collection of Lie algebra elements $X_w$ constant on each wedge.
Moreover, as we shall see, the integration over group elements
$g_\epsilon$ in (\ref{k1}) renders all $X_w$ to have the same
norm in each disjoint component of $\cal M$, which is in agreement
with what we get in the canonical approach.

The group elements $g_\epsilon$ in (\ref{k1}) are ``shared''
by two dual faces. Thus, all integrals over $g_\epsilon$
in (\ref{k1}) have as the integrand the product of two matrix
elements of $g_\epsilon$. These can be taken using the formula
(\ref{a1}) of the Appendix. To describe the result of this 
integration we will assume that $\cal M$ has a single connected
component. Then the generating functional in two dimensions 
is given by the following expression
\begin{equation}\label{z2d}
Z_2(J,\Delta)=\sum_j ({\rm dim}_j)^{V-E} \prod_f 
P(J_1) P(J_2) P(J_3)\,                         
\raisebox{-35pt}{\epsfig{figure=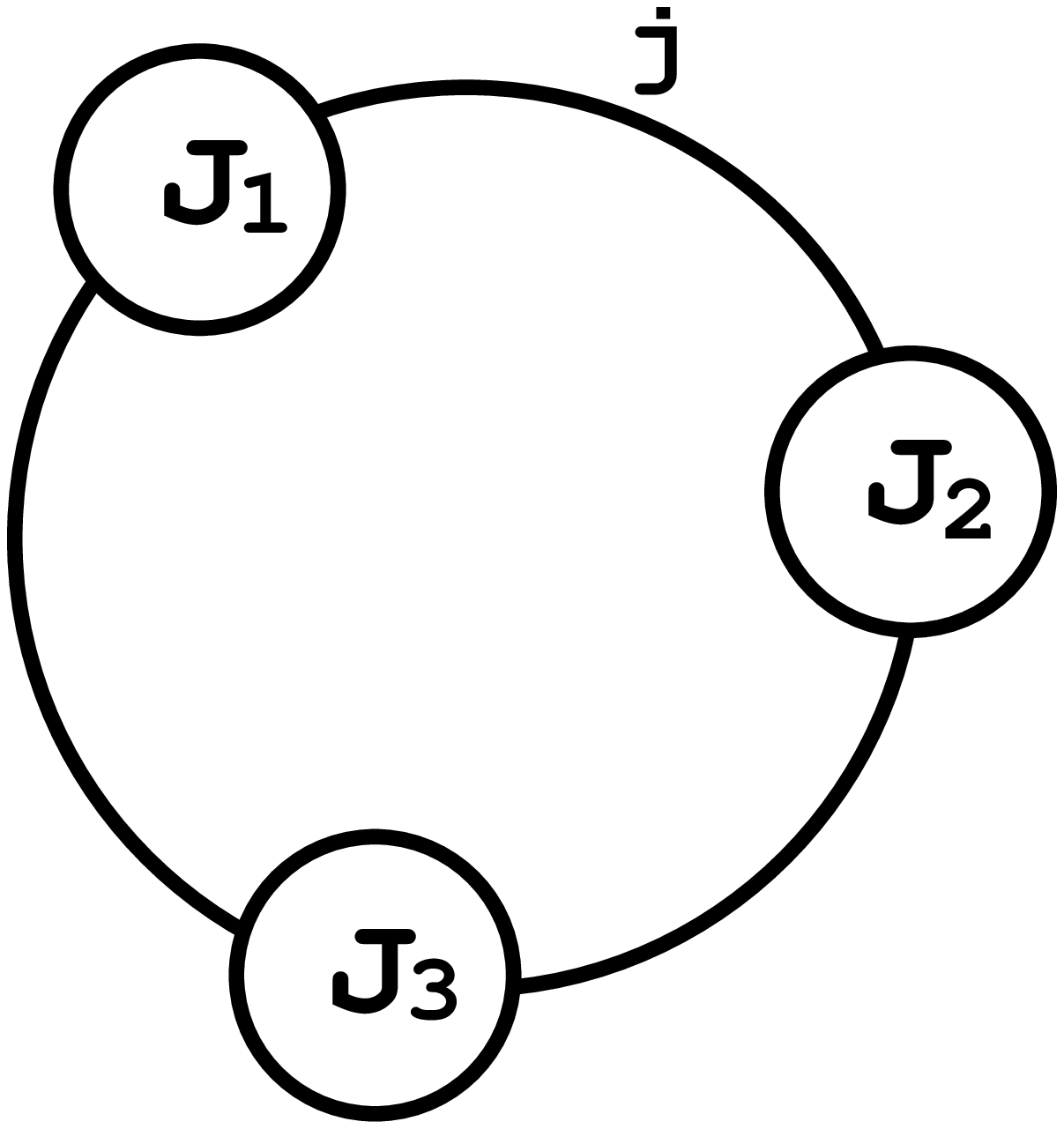,height=70pt}} 
\end{equation}
In the integration over the group elements $g_\epsilon$ in (\ref{k1})
survive only the terms in which all spins $j_\sigma$ are equal.
This spin is denoted by $j$ in (\ref{z2d}). The symbols $V,E$
stand in (\ref{z2d}) for number of vertices and number of edges 
in the triangulation correspondingly. The product here is taken over
all centers of faces of $\Delta$, or, equivalently, over faces $f$. 
The graphical notation stands for 
\begin{equation}
\raisebox{-35pt}{\epsfig{figure=2d.eps,height=70pt}}
= \chi_j(\exp{J_1}\exp{J_2}\exp{J_3}),
\end{equation}
where $J_1,J_2,J_3$ are the three Lie algebra elements (currents)
``shared'' by the face $f$. Recall that in two
dimensions each face is just a union of three wedges, and 
$J_1,J_2,J_3$ are the currents $J_w$ corresponding to these wedges
(see Fig. 6).
\begin{figure}\label{gr1}
\centerline{\epsfig{figure=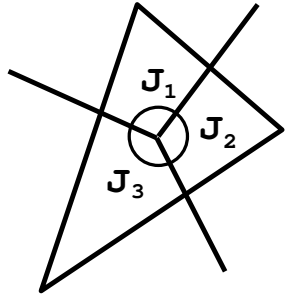,height=1.5in}}
\bigskip
\caption{In two dimensions each face is a union of three wedges, and 
$J_1,J_2,J_3$ are the currents $J_w$ corresponding to these wedges.}
\end{figure}
The expression (\ref{z2d}) is the final result for the generating
functional in two dimensions.

Note that the graph in this graphical representation of the
character can be obtained as a result of the following
simple construction. Let us draw a circle $S^1$ centered at 
the center of the face (see Fig. 6). The intersection
of wedges with the circle gives the circle itself. The
intersection of this circle with the dual edges gives three
points on the circle. Each wedge is labelled by a spin $j$,
which is the same for all wedges, and we can assign a label
$j$ to the circle that intersects the three wedges.
Then the face contribution to (\ref{z2d}),
that is, the contribution that we graphically represent
by (\ref{gr1}), is simply the {\it spin network} constructed
with the labelled circle, with insertion of the
three group elements $(\exp{J_1}\exp{J_2}\exp{J_3})$
at the points where the circle is intersected by the
corresponding wedges. This construction is trivial in two
dimensions, but turns out to be generalizable to
any dimension.

\subsection{3 dimensions}
\label{sec:3d}

The three-dimensional case is more interesting. Here the dual 
faces no longer cover the spacetime manifold, and the $E$ field
acquires a true distributional character.  The general construction 
prescribes to replace the $E$ field, which is now a one-form, by a 
distribution concentrated along dual faces and constant along the 
wedges.  It is illustrative to give a coordinate expression for such a 
distributional field.  Let us consider an arbitrary wedge $w$ (see 
Fig.  7).
\begin{figure}
\centerline{\epsfig{figure=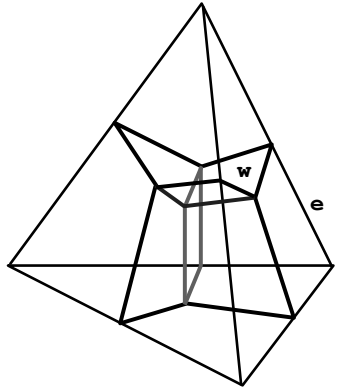,height=2in}}
\bigskip
\caption{Tetrahedron from a triangulation of a 3-dimensional spacetime
manifold. The figure also shows the wedges that lie inside this 
tetrahedron.}
\end{figure}
Let $u=0$ be the equation of a plane containing $w$. Then the field $E$ that
has the correct distributional character is given by the
expression
\begin{equation}\label{distr-2}
du\,\delta(u)\,X_w.
\end{equation}

To calculate the generating functional let us note that each edge of
the dual triangulation belongs to three dual faces. Thus, in (\ref{k1}),
one has to take integrals over products of three matrix elements.
The corresponding formula is given in the Appendix \ref{appHa}.  The 
result of this integration can be described as a generalization of the 
``circle'' construction given at the end of the previous subsection.  
Let us draw a sphere $S^2$ centered at the center of each tetrahedron.  
The wedges belonging to a particular tetrahedron intersect the sphere 
and draw a graph on its surface.  We will refer to this graph as 
$\Gamma_t$, where $t$ refers to a tetrahedron of the triangulation 
that was used to construct $\Gamma_t$.  It is not hard to see that 
$\Gamma_t$ is a tetrahedron whose vertices come from the intersection 
of the sphere with the dual edges.  The same graph $\Gamma_t$ can also 
be obtained by looking at the boundary of tetrahedron $t$ (this 
boundary has the topology of $S^2$), which is triangulated by the 
faces of $t$, and constructing the graph dual to the triangulation of 
this $S^2$.  The resulting tetrahedron is the same as that in the 
previous construction with a sphere.

In (\ref{k1}) the sum is taken over spins $j_\sigma$ labelling
the dual faces. The result of the integration over the 
group elements $g_\epsilon$ will still be a sum over spins
$j_\sigma$. Each wedge belongs to some dual face, and, thus,
is labelled by spin.  As we have just seen the edges of $\Gamma_t$ are in 
one-to-one correspondence with the wedges.  Let us label these 
edges with the same spins as those of the corresponding wedges.  We 
have a current $J_w$ associated with each of the 6 wedges belonging to 
tetrahedron $t$.  Let us denote this currents by $J_1,\ldots,J_6$.  We 
can now construct a function of the group elements 
$\exp{J_1},\ldots,\exp{J_6}$ that is just a spin network function.

The generating functional in three dimensions is then given 
by the following expression
\begin{equation}\label{z3}
Z_3(J,\Delta)=\sum_{j_\sigma} 
\prod_\sigma {\rm dim}_{j_\sigma} \prod_t 
P(J_1)\cdots P(J_6)
\raisebox{-50pt}{\epsfig{figure=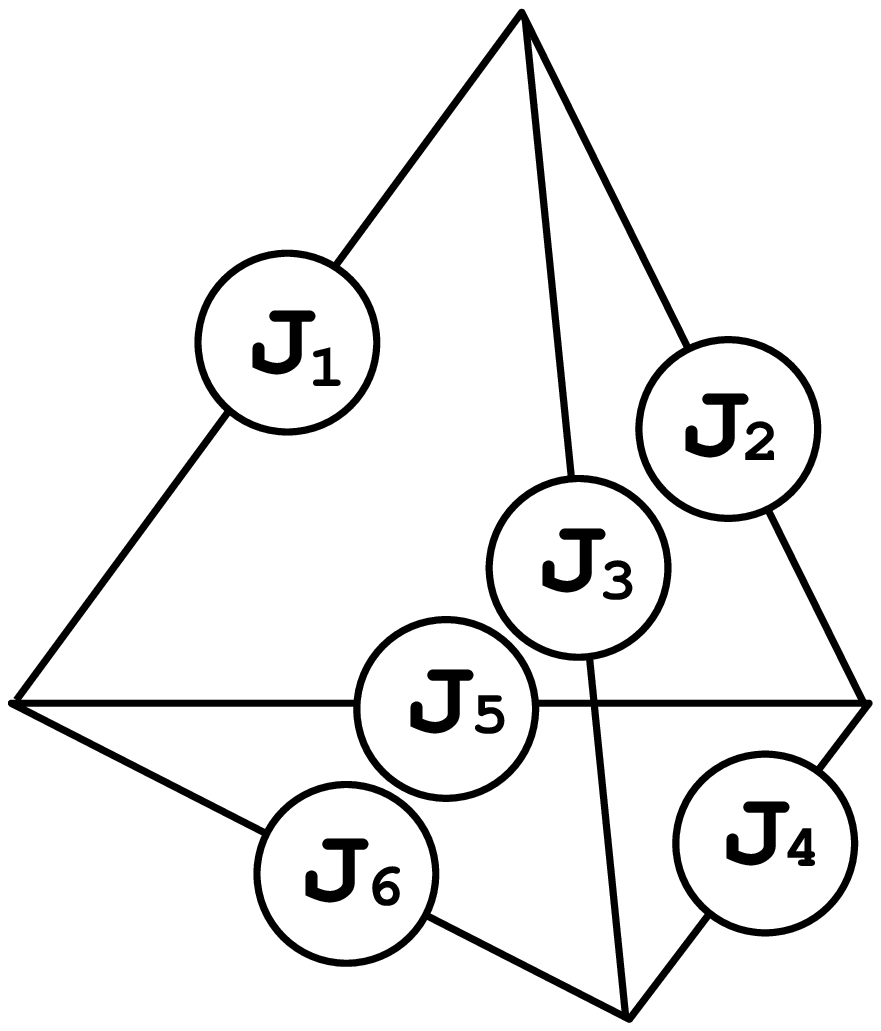,height=100pt}} ,
\end{equation}
where the sum is over all possible coloring of dual faces. This is the final 
expression we are going to use in the following section  in order to 
derive spin foam models.

\subsection{4 dimensions}
\label{sec:4d}

The case of four dimensions is analogous to the just
analyzed case of three dimensions. The only difference
is that it is harder to visualize a four-dimensional
triangulated manifold, and that the final expression
for the generating functional is more complicated.
However, the final result for the generating 
functional follows the same pattern.

First, let us give a coordinate expression for the
distributional field $E$ in four dimensions.
Let $u,v$ be two functions 
such that $u=0,v=0$ is the equation for a 2-surface in $\M$ containing
one of the faces $\sigma$ of the dual complex, and $du\wedge dv$ is positive
as defined by the orientation of $\sigma$. 
We then define $E$ to be equal to 
\begin{equation} \label{distr:4d}
du\wedge dv\,\delta(u)\delta(v)\,X_w
\end{equation}
on each wedge $w$, and to be equal to zero everywhere else.
We repeat this construction for all wedges, and add all these
distributional forms together to get a two-form that is
concentrated along the faces of the dual complex. Then the integral
of such distributional $E$ over a wedge of the triangulation
is equal simply to $X_w$ corresponding to that wedge.

Let us now describe the result of integration over the group 
elements $g_\epsilon$ in (\ref{k1}). In four dimensions each 
dual edge is shared by four dual faces. Thus, one has to
take the integral of the product of four matrix
elements.  The required formula is given in Appendix~\ref{appHa}.  The 
result of this integration can again be described using a certain spin 
network function of the group elements corresponding to the currents 
$J_w$.  As in the case of two and three dimensions, let us introduce 
special graphs in the vicinity of the center of each 4-simplex $h$, 
which we will call $\Gamma_h$.  We define graph $\Gamma_h$ as the 
intersection of the sphere $S^3$ surrounding the center of 4-simplex 
$h$ with the dual faces $\sigma$.  The graph $\Gamma_h$ lives in 
$S^3$, but it can be projected on a plane.  This gives us the 
pentagon graph.  Note that at this stage we do not care about types 
of crossings we get (that is, whether this is under- or 
over-crossing).  The edges of this graph are in one-to-one 
correspondence with the wedges $w$ belonging to the 4-simplex $h$.  
Thus, we can associate to each edge a current $J_w$, and label it with 
the spin $j_\sigma$ labelling the dual face to which the corresponding 
wedge belongs.  Additionally, let us label each of five vertices of 
the pentagon by a half-integer (spin).  Vertices of the pentagon are 
in one-to-one correspondence with tetrahedra of the triangulation.  
Thus, we shall use the notation $j_t$ for this spins.

Inside each four-simplex there are 10 wedges. Let
us denote the corresponding currents by $J_1,\ldots,J_{10}$.
One can then construct the function of
the group elements $\exp{J_1},\ldots,\exp{J_{10}}$.

The generating
functional $Z_4$ is given by
the sum over spins of
products over 4-simplices of the above functions
of the currents $J_w$
\begin{equation}\label{z4d}
Z_4(J,\Delta) = \sum_{j_\sigma, j_t} 
\prod_\sigma {\rm dim}_{j_\sigma} 
\prod_t {\rm dim}_{j_t} \prod_h 
P(J_1)\cdots P(J_{10})
\raisebox{-50pt}{\epsfig{figure=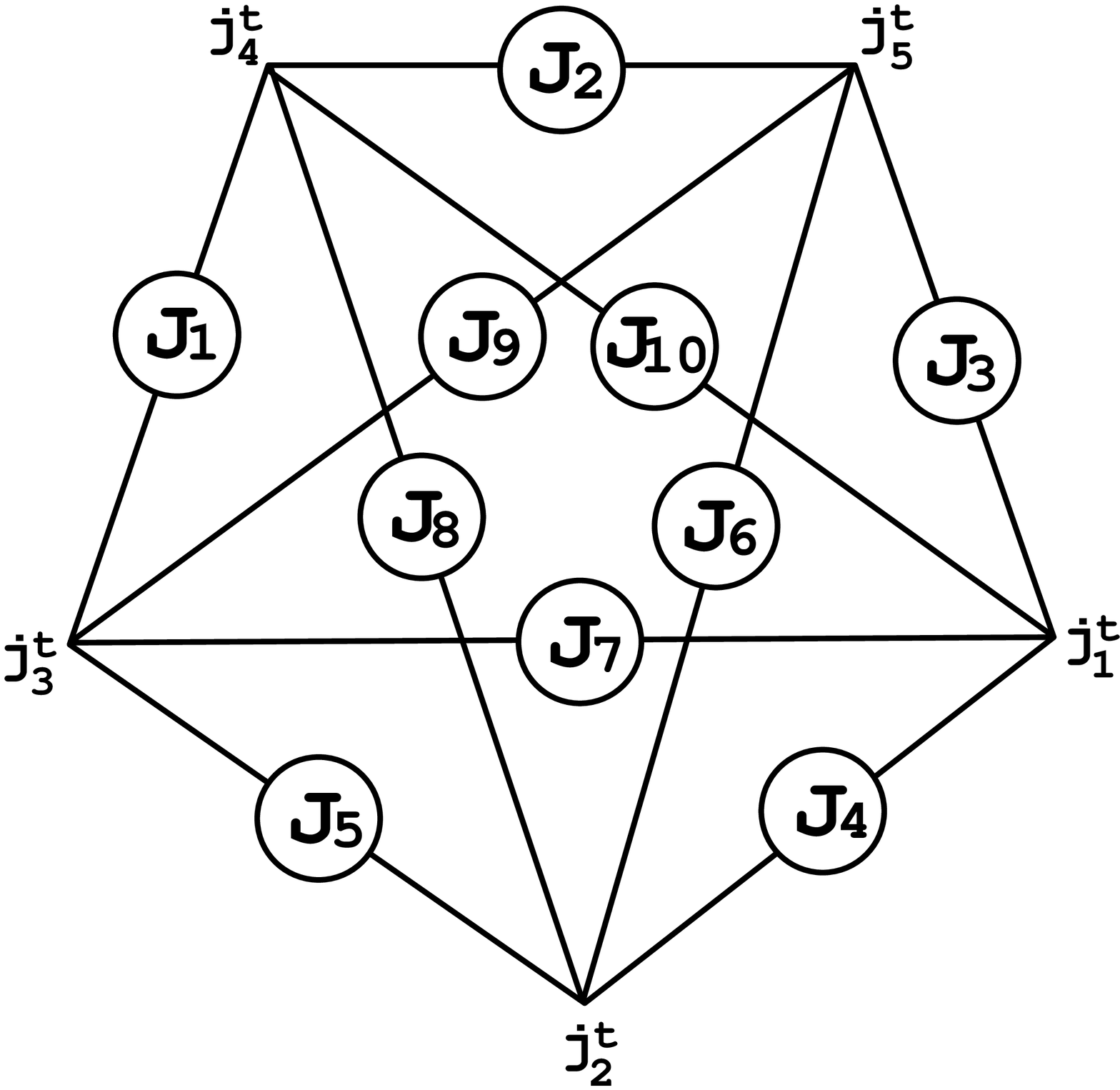,height=100pt}} 
\end{equation}
Here $h$ labels 4-simplices of the triangulation and
$j_\sigma$ are spins labelling the dual faces. The
resolution of each vertex in this pentagon graph is given
in the Appendix \ref{appHa}.

\section{Theories of interest}
\label{sec:th}

Having calculated the generating functional in various spacetime
dimensions we are in the position to use our results to study
concrete theories. However, let us first specify in more details
the class of theories that can be treated within our approach, and
give physically interesting examples of theories belonging to this class.
In this paper we will restrict our
attention to a special class of theories.  For simplicity we
fix the  gauge group to be $G=\SU(2)$.  Let $\M$ be an oriented smooth 
d-dimensional manifold, and let $P$ be an $\SU(2)$-bundle over $\M$.  
The basic fields of a theory of interest are a connection $A$ on $P$, 
and an ${\rm Ad} P$-valued (d-2)-form, often called $B$, which we call 
$E$ because of its relation with the $\SU(2)$ ``electric'' field of 
the canonical formulation of the theory.  The action of the theory is 
of the form
\begin{equation} \label{type}
S[A,E]=\int_\M \left[ {\rm Tr}\left( E\wedge F \right) + \Phi(E)\right],
\end{equation}
where $F$ is the curvature of $A$, `Tr' is the trace in the
fundamental representation of $\SU(2)$, and $\Phi$ is a gauge 
invariant polynomial
function that depends on $E$ but not on its derivatives. 
Also, $\Phi(E)$ should be 
a d-form so that it can be integrated. 
It can also depend on other dynamical 
fields, for example Lagrange multipliers as in the case of gravity,
or on additional non-dynamical fields (background structure),
as in the case of Yang-Mills theories.
Let us now give examples of theories belonging to this class.

\subsection{BF theories}

The action of BF theory in any dimension is given by
\begin{equation}\label{bfaction}
\int_{\cal M} {\rm Tr}(E\wedge F),
\end{equation}
that is, this is the action of the type (\ref{type}) with 
zero ``interaction'' term $\Phi$. The equations of motion
that follow from this action state that
\begin{equation}
d_A E = 0, \qquad\qquad F=0,
\end{equation}
where $d_A E$ is the covariant derivative of $E$.
Thus, the flat connections play the dominant role in BF theory.

\subsection{2D Yang-Mills}
\label{sec:ymcl}

Yang-Mills action can be written only if one
introduces a background metric on $\cal M$. However, in 
two spacetime dimensions, Yang-Mills action turns out 
to depend only on a measure (area two-form) defined by
the metric on $\cal M$.  Thus, we will keep track only of this 
dependence of the action on a measure $d\mu$ on $\cal M$.  Yang-Mills 
theory in 2D is described by the following BF-like action
\begin{equation}\label{ym2}
\int_{\cal M} {\rm Tr}(E F) +\,{e^2\over 2} \int_{\cal M} d\mu 
{\rm Tr} (E^2).
\end{equation}
Here $E$ is a Lie-algebra valued zero-form, $F$ is the curvature
of the connection $A$. Solving equations of motion for
$E$ that follow from this action, and substituting the 
solution into the action, one can check that the action
reduces to the standard Yang-Mills action
\begin{equation}\label{ymaction}
S_{\rm YM} = - {1\over 4e^2} \int_{\cal M} d^2x \sqrt{g} g^{ac} g^{bd}
{\rm Tr}(F_{ab}F_{cd}),
\end{equation}
where $g_{ab}$ is a
metric (of Euclidean signature) on $\cal M$ such that
$ d^2x \sqrt{g} = d\mu$,
where $\sqrt{g}$ is the square root of the determinant 
of the metric.

Using the techniques developed in this paper, one can calculate 
the vacuum-vacuum transition amplitude of this 
theory, that is the path integral of $\exp{(i\times{\rm action})}$.  
However, an interesting feature of this theory is that the same 
techniques can be used to calculate the {\it partition function}, that 
is, the path integral
\begin{equation}\label{ym1}
\int {\cal D}A {\cal D}E \exp\left(-S_{\rm YM}\right).
\end{equation}
The problem of calculation of the partition function
can be reduced to the problem of calculation of the vacuum-vacuum
transition amplitude of a somewhat different theory, that is 
of the path integral of $\exp{(i\times{\rm action})}$, with the
action given by
\begin{equation}\label{ym3}
\int_{\cal M} {\rm Tr}(E F) - i\,{e^2\over 2} 
\int_{\cal M} d\mu {\rm Tr} (E^2).
\end{equation}
Indeed, integrating over $E$ in the path integral of 
$\exp{(i\times{\rm action})}$, one gets (\ref{ym1}). As one
can see, the two actions (\ref{ym2}),(\ref{ym3}) differ 
by the factor of $i$ in front of the second term. Using the 
two actions (\ref{ym2}),(\ref{ym3}) we will calculate below both the 
vacuum-vacuum transition
amplitude and partition function of the theory.

\subsection{3D BF theory with the `cosmological term'}

In three dimensions the $E$ field of BF theory is
a one-form, and one can add a term  cubic in $E$
to the BF action to obtain
\begin{equation}\label{bf3daction}
-\int_{\cal M} {\rm Tr}\left(E\wedge F + 
{\Lambda\over 12} E\wedge E\wedge E\right),
\end{equation}
where $\cal M$ is assumed to be a three-dimensional orientable
manifold. The action is a functional of an $\SU(2)$ connection $A$,
whose curvature form is denoted by $F$,
and a 1-form $E$, which takes values in the Lie algebra of
$\SU(2)$. Thus, the action (\ref{bf3daction}) is that of BF
theory in 3d, with $E$ field playing the role of $B$, and with
an additional ``cosmological term'' added to the usual BF action.
This theory is related to gravity in 3D as follows.
Having the one-form $E$, one can construct from it
a real metric of Euclidean signature 
\begin{equation}\label{metric}
g_{ab} = - {1\over 2}{\rm Tr}(E_a E_b).
\end{equation}
Here $a,b$ stand for spacetime indices: $a,b,...=1,2,3$.
We take the $E$ field to be anti-hermitian, which 
explains the minus sign in (\ref{metric}).
Thus, the $E$ field in (\ref{bf3daction}) has the interpretation
of the triad field. One of the equations of 
motion that follows from (\ref{bf3daction}) states that $A$
is the spin connection compatible with the triad $E$. 
Taking the $E$ field to be non-degenerate and ``right-handed'', 
i.e., giving a  positive-definite volume form 
\begin{equation}\label{vform}
{1\over 12} \tilde{\varepsilon}^{abc}
{\rm Tr}(E_a E_b E_c),
\end{equation}
and substituting into (\ref{bf3daction}) the
spin connection instead of $A$, 
one gets the Euclidean Einstein-Hilbert action
\begin{equation}
{1\over2}\int_{\cal M} d^3x \sqrt{g}\,(R-2\Lambda).
\end{equation}
We use units in which $8\pi G=1$. The minus in front
of (\ref{bf3daction}) is needed to yield precisely the Einstein-Hilbert action
after the elimination of $A$. 
Thus, $\Lambda$ in (\ref{bf3daction}) is
the cosmological constant. An important difference
of this theory and Einstein's theory is the fact BF
theory, unlike gravity, is defined for degenerate
metrics.

\subsection{4D BF theory with the `cosmological term'}

In four spacetime dimensions the $E$ field of BF theory is a
two-form. Thus, one can construct the following action
functional
\begin{equation}\label{bf4daction}
\int_{\cal M} {\rm Tr}(E\wedge F) - {\Lambda\over 2} {\rm Tr}(E\wedge E),
\end{equation}
where $\Lambda$ is a real parameter, which we will refer to as the
``cosmological constant''. If one, as in the following subsection, 
adds to this action an additional constraint that $E$ is simple,
that is, given by a product of two one-forms, then this theory
is equivalent to Einstein's theory and $\Lambda$ is proportional
to the ``physical'' cosmological constant.

\subsection{4D Euclidean self-dual gravity}

The action for Euclidean general relativity in the self-dual 
first order formalism is given by \cite{Actions}
\begin{equation} \label{sdgraction}
\int_{\cal M} {\rm Tr}(E\wedge F) - \psi^{ij}
\left(E^i\wedge E^j - {1\over 3}\delta^{ij} E^k\wedge E_k\right),
\end{equation}
where, to write $E^i$, we have introduced a basis in the
Lie algebra of $\SU(2)$. Here $\psi^{ij}$ is a symmetric matrix
of Lagrange multipliers. The variation of the action with 
respect to $\psi$ yields equations 
\begin{equation}\label{cons}
E^{i}\wedge E^{j} = \delta^{ij} \frac{1}{3} E^{k}\wedge E_{k}.
\end{equation}
In the case $E$ is non degenerate, i.e., the right hand side of (\ref{cons})
is non zero, these equations are satisfied if and only if $E$ is the 
self-dual part of a decomposable 2-form, i.e., if and only if there 
exists a tetrad field $e^{I},\, I=0,1,2,3$ such that $E^{i} = \pm ( 
e^{0}\wedge e^{i} + \frac{1}{2} \epsilon_{ijk} e^{{j}}\wedge e^{k})$.  
In this case the action reduces to the self-dual Hilbert-Palatini 
action for 4-dimensional gravity and $\psi_{ij}$ correspond to the 
components of the self-dual part of the Weyl curvature tensor.  Thus, 
(\ref{sdgraction}) is the BF action in four dimensions, with an 
additional ``simplicity'' constraint added to it. 

\section{Corresponding spin foam models}
\label{sec:sf}

In this section we present state sum models for the theories we 
just described.

\subsection{Ponzano-Regge-Ooguri models}

These models give an expression for the vacuum-vacuum transition
amplitude of BF theory, i.e., the path integral
\begin{equation}\label{zbf}
Z^{{\rm BF}}({\cal M}) = \int {\cal D}A {\cal D}E 
\exp\left(i\int_{\cal M}{\rm Tr}(E\wedge F) \right),
\end{equation}
as a sum over colored triangulations. More precisely, let
us fix a triangulation $\Delta$ of $\cal M$. In three
dimensions, let us label the edges of $\Delta$ by 
irreducible representations of $\SU(2)$, that is,
by spins $j$. Thus, to each edge $e$ we assign a label
$j_e$. One can then construct the following sum
over labellings
\begin{equation}\label{pr}
{\rm PR}(\Delta) = \sum_{j_e} \prod_e 
{\rm dim}_{j_e} \prod_t (6j)_t.
\end{equation}
Here ${\rm dim}(j)=2j+1$ is the dimension of the 
representation $j$, the second product is taken over
tetrahedra $t$ of $\Delta$, and $(6j)$ is the (normalized) classical
(6j)-symbol (see the Appendix C for a definition) 
constructed from the six spins 
labelling the edges of $t$.  Note that 
the sum over spins in (\ref{pr}) diverges. To make
sense of it one must introduce a regularization.
A possible regularization is given by the Turaev-Viro
model, which we discuss below. After the introduction of
this  ``regularization'', this sum 
can be shown to be triangulation independent.  Thus, (\ref{pr}) gives 
an invariant of $\cal M$: ${\rm PR}(\Delta)={\rm PR}(\cal M)$.

In four dimensions one can construct a similar sum
by labelling the faces of $\Delta$. Let
us denote the spin labelling a face $f$ by $j_f$.
Let us consider the following quantity:
\begin{equation}\label{o}
{\rm O}(\Delta) = \sum_{j_f,j_t} \prod_f 
{\rm dim}_{j_f} \prod_t {\rm dim}_{j_t}
\prod_h (15j)_h,
\end{equation}
After a certain ``regularization'' 
(given by the Crane-Yetter model below) that gives
sense to the infinite sums in (\ref{o}), this
can be shown to be triangulation independent: 
${\rm O}(\Delta)={\rm O}(\cal M)$. In (\ref{o}) 
one has introduced an additional label $j_t$ for
each tetrahedron of $\Delta$. The spin $j_t$
labels an intertwiner one has to assign to each 
tetrahedron (see the subsection on Crane-Yetter model
for details). The last product
is taken over the 4-simplices $h$ of $\Delta$. Then
$(15j)$ is the (normalized) (15j)-symbol constructed from 
ten spins $j_f$ labelling the faces of $h$ and five spins
$j_t$ labelling the five tetrahedra composing $h$. See
the Appendix \ref{appHa} for a definition of the normalized (15j)-symbol.

As was realized by Ooguri \cite{Ooguri}, these models
give a ``discrete'' realization of the path
integral because they can be obtained from the 
requirement that the connection on $\cal M$ is
flat. Indeed, formally the path integral
(\ref{zbf}) is equal to 
\begin{equation}
\int {\cal D}A \delta(F).
\end{equation}
The state sums (\ref{pr}),(\ref{o}) are in certain
precise sense realizations of the integral over
connections with the integrand being the delta-function at $F=0$.

\subsection{Migdal-Witten model}

Migdal \cite{Migdal} has studied 
the Yang-Mills theory on a lattice, and, in particular, has 
proposed a lattice model for the Yang-Mills theory in two 
dimensions. This model was later studied by Witten \cite{Witten}
in the connection with topological field theories in two dimensions.

As we discussed above, the problem of calculation of the Yang-Mills 
partition function can be reduced
to the problem of calculation of the path integral
\begin{equation}\label{ympart}
Z^{\rm YM}(\rho e^2,{\cal M}) = 
\int {\cal D}A {\cal D}E \exp\left(i\int_{\cal M}{\rm Tr}(E F) +
{e^2\over 2}\int_{\cal M} d\mu {\rm Tr}(E^2)\right),
\end{equation}
where $\rho$ is the total area of $\cal M$, and other
notations are explained in subsection  (\ref{sec:ymcl}). Migdal
\cite{Migdal} has proposed the following lattice version of this
path integral. For simplicity, we will formulate Migdal model
on a triangulated manifold. Let us triangulate our
two-dimensional manifold $\cal M$, and introduce the
dual triangulation (see Fig. 5). Let us label the edges $\epsilon$ of the 
dual complex by group elements $g_\epsilon$, and the dual faces $\sigma$ by
irreducible representations of $\SU(2)$, i.e., spins $j_\sigma$.
The ``discrete'' version of (\ref{ympart}) is then given by
(see \cite{Witten}):
\begin{equation}\label{ym}
{\rm YM}(\rho e^2,{\cal M}) = \prod_\epsilon \int dg_\epsilon 
\sum_{j_\sigma} \prod_\sigma
{\rm dim}_{j_\sigma} 
\chi_{j_\sigma}(g_{\epsilon_1}\cdots g_{\epsilon_n})
\exp\left(-e^2\rho_\sigma c(j_\sigma)/2\right),
\end{equation}
where the multiple integral is performed over all group
elements $g_\epsilon$ and $dg$ is the normalized Haar measure
on $\SU(2)$, $\rho_\sigma$ is the area of the dual face
$\sigma$, as defined by the measure $d\mu$, $c(j)$ is the
value of the quadratic Casimir operator in the representation
$j$
\begin{equation}
c(j) =  2j(j+1),
\end{equation} 
and $g_{\epsilon_1}\cdots g_{\epsilon_n}$ is the product
of group elements around the dual face.
After integration over the group elements $g_\epsilon$ 
the partition function takes the following simple form:
\begin{equation}\label{ymm2}
\sum_j ({\rm dim}_j)^{\kappa(\cal M)} e^{-e^2\rho c(j)},
\end{equation}
where $\kappa(\cal M)$ is the Euler characteristics of $\cal M$.
Thus, as we indicated in the 
argument of YM, the partition function 
depends only on the topological properties of $\cal 
M$.  Also, the dependence on measure $d\mu$ enters only through the 
dependence on the total area $\rho$ of $\cal M$.

\subsection{Turaev-Viro model}
\label{3C}

The Turaev-Viro model gives a way to calculate
the transition amplitude of the theory
defined by (\ref{bf3daction}), i.e., the path integral
\begin{equation}\label{zbfl3d}
Z^{\rm BF}_3(\Lambda,{\cal M}) = \int {\cal D}E \,{\cal D}A \, 
\exp{\left( -i \int_{\cal M} {\rm Tr}\left(E\wedge F + 
{\Lambda\over 12} E\wedge E\wedge E\right)\right)}.
\end{equation}
In this paper we consider only vacuum-vacuum amplitudes. Although
the Turaev-Viro model can be used to calculate more general
amplitudes between non-trivial initial and final states, we
will not use this aspect of the model here. 
We consider the version of the model formulated on a triangulated 
manifold. Thus, let us fix a triangulation $\Delta$ of $\cal M$. 
Let us label the edges, for which we will employ the 
notation $e$, by irreducible representations of the
quantum group $(\SU(2))_q$, where $q$ is a root of unity
\begin{equation}\label{q}
q = e^{{2\pi i\over k}} \equiv e^{i\hbar}.
\end{equation} 
Later we will relate the parameter $\hbar$ with 
the cosmological constant $\Lambda$. The irreducible representations
of $(\SU(2))_q$ are labelled by half-integers (spins) $j$ satisfying
$j \leq (k-2)/2$. Thus, we associate a spin $j_e$ to each 
edge $e$. The vacuum-vacuum transition amplitude of the theory is then
given by the following expression (see, for example, \cite{Roberts}):
\begin{equation}\label{tv}
{\rm TV}(q,\Delta) = \eta^{2V}\,\sum_{j_e} \prod_e {\rm dim}_q(j_e) 
\prod_t (6j)_q,
\end{equation}
where $\eta$ and the so-called quantum dimension ${\rm dim}_q(j)$
are defined in the Appendix A by (\ref{eta}) and (\ref{qdim}) 
correspondingly, and $V$ is the number of vertices in $\Delta$.  The 
last product in (\ref{tv}) is taken over tetrahedra $t$ of $\Delta$, 
and $(6j)_q$ is the (normalized) quantum $(6j)$-symbol constructed 
from the 6 spins labelling the edges of $t$ (see Appendix C).  It 
turns out that (\ref{tv}) is independent of the triangulation $\Delta$ 
and gives a topological invariant of $\cal M$: ${\rm 
TV}(q,\Delta)={\rm TV}(q,\cal M)$.

The construction that interprets the Turaev-Viro invariant (\ref{tv}) 
as the vacuum-vacuum transition amplitude of the theory defined
by (\ref{bf3daction}) is as follows. It has been proved (see {\it e.g.}
\cite{Roberts}) that (\ref{tv}) is equal
to the squared absolute value of the Chern-Simons amplitude
\begin{equation}\label{tv-cs}
{\rm TV}(q,{\cal M}) = |{\rm CS}(k,{\cal M})|^2,
\end{equation}
with the level of Chern-Simons theory being equal to $k$
from (\ref{q}). It is known, however, that the action
(\ref{bf3daction}) can be written as a difference of two copies of
Chern-Simons action
\begin{equation}\label{csaction}
S_{\rm CS}(A) = {k\over 4\pi} \int_{\cal M}
{\rm Tr}\left( A\wedge dA + {2\over 3} A\wedge A\wedge A\right).
\end{equation} 
Indeed, note that
\begin{equation}\label{two}
S(A+\lambda E) - S(A-\lambda E) = {k\lambda\over\pi} \int_{\cal M} 
{\rm Tr}\left( E\wedge F + {\lambda^2\over 3} E\wedge E\wedge E\right),
\end{equation}
where $\lambda$ is a real parameter. Thus, (\ref{two}) is
equal to (\ref{bf3daction}) if 
\begin{equation}\label{rel}
\lambda = - \sqrt{\Lambda}/2, \qquad k = {2\pi\over\sqrt{\Lambda}},
\qquad {\rm or} \qquad \hbar = \sqrt{\Lambda}.
\end{equation}
This relates the deformation parameter $q$ of the Turaev-Viro
model with the cosmological constant $\Lambda$, and proves
that the Turaev-Viro amplitude is proportional to the 
vacuum-vacuum transition amplitude of the theory defined by
(\ref{bf3daction})
\begin{equation}
{\rm TV}(q,{\cal M}) \propto Z^{\rm BF}_3(\Lambda,{\cal M}).
\end{equation}

To compare in Sec. \ref{sec:appl} the spin foam model
obtained via our techniques with the Turaev-Viro 
model, we will need the first-order term in the
decomposition of (\ref{tv}) in the power series in
$\Lambda$. This is proportional to 
the expectation value of the spacetime volume in BF theory
at $\Lambda=0$. Indeed, the expectation value of 
the volume is given simply by the derivative of
the amplitude (\ref{z}) with respect to $(-i\Lambda)$
evaluated at $\Lambda=0$:
\begin{eqnarray}\label{der}
\langle{\rm Vol}\rangle = 
{\int {\cal D}E {\cal D}A \, {\rm Vol}({\cal M}) e^{iS}
\over \int {\cal D}E {\cal D}A \, e^{iS}} = 
i\,\left({\partial \ln{Z^{\rm BF}_3(\Lambda)}
\over\partial \Lambda}\right)_{\Lambda=0}, \\
{\rm Vol}({\cal M}) = \int_{\cal M}
{1\over 12} \tilde{\varepsilon}^{abc}
{\rm Tr}(E_a E_b E_c).
\nonumber
\end{eqnarray}
Thus, the expectation value of the volume of $\cal M$
in BF theory is proportional to the first order
term in $\Lambda$ in the decomposition of (\ref{tv}).

Let us find this expectation value.
An important subtlety arises here. It is not hard to show
that (\ref{tv}) has the following asymptotic expansion
in $\hbar$
\begin{equation}\label{dec}
\left({\hbar^3\over 4\pi}\right)^V\,{\rm PR}(\Delta)\,
\left(1-\hbar^2 i\langle{\rm Vol}\rangle\right),
\end{equation}
where $\rm PR$ is the amplitude of the Ponzano-Regge model 
(\ref{pr}) $V$ is the number of vertices in $\Delta$,
and $i\langle{\rm Vol}\rangle$ is a real quantity independent of $\hbar$.
Thus, apparently there is no term proportional to $\hbar^2=\Lambda$
in this expansion. However, as it is explained, for
example, in \cite{Volume},
the state sum invariant (\ref{tv}) does not exactly give 
the transition amplitude of BF theory. Instead, the 
Turaev-Viro amplitude is only proportional to 
the BF amplitude, and the proportionality coefficient
depends on $\hbar$. This can be understood as follows.
The integration over $(A+\lambda E), (A-\lambda E)$,
which is carried out to obtain $|{\rm CS}(k,{\cal M})|^2$ in (\ref{tv-cs})
and thus the Turaev-Viro amplitude, 
is different from the integration over $A,E$
one has to perform to obtain (\ref{zbfl3d}). The difference in the 
integration measures is a power of $\hbar$. Thus, the 
amplitude (\ref{zbfl3d}) and the squared absolute value of the 
amplitude of the Chern-Simons theory are proportional to
each other with the coefficient of proportionality being
a power of $\hbar$. In the discretized version
of the theory, given by the Turaev-Viro model, this
power of $\hbar$ is replaced by $\hbar^{3V}$.
Thus, the Turaev-Viro amplitude, in the
limit of small cosmological constant, differs from the
BF amplitude by a power of $\hbar^{3V}$. 

This remark being made, we can write an expression for
the expectation value of the volume in BF theory:
\begin{eqnarray}\nonumber
i\langle{\rm Vol}\rangle_\Delta = - {\partial\over\partial\Lambda} 
\left( {{\rm TV}(\Lambda,\Delta)\over {\rm PR}(\Delta)(\hbar^3/4\pi)^V} 
\right)_{\Lambda=0} = \\
{1\over {\rm PR}(\Delta)} \sum_{j_e} i{\rm Vol}(\Delta,{\bf j}) 
\left( \prod_e {\rm dim}(j_e) \prod_t (6j)\right),
\label{vol-exp}
\end{eqnarray}
where the function ${\rm Vol}(\Delta,{\bf j})$ of the triangulation $\Delta$
and the labels ${\bf j}=\{j_e\}$ is given by
\begin{eqnarray}
i{\rm Vol}(\Delta,{\bf j}) &=& \nonumber
\sum_v \left( - {\partial\over\partial\Lambda}
\left( {\eta^2 \over (\hbar^3/4\pi)} 
\right)\right)_{\Lambda=0} +
\sum_e \left(- {\partial\ln({\rm dim}_q(j_e))\over\partial\Lambda}
\right)_{\Lambda=0} \\ &&+
\sum_t \left(- {\partial\ln((6j)_q)\over\partial\Lambda}
\right)_{\Lambda=0}. \label{vol-1}
\end{eqnarray}
Here $v$ stands for vertices of $\Delta$, $e$ stands for edges and
$t$ stands for tetrahedra.
We intentionally wrote the expectation value of the volume
in the form (\ref{vol-exp}) to introduce the volume ${\rm Vol}(\Delta,{\bf j})$
of a labelled triangulation, which will be of interest to us
in what follows. Indeed, (\ref{vol-exp}) has the form 
\begin{equation}
{\sum_{j_e} i{\rm Vol}(\Delta,{\bf j}) {\rm Amplitude}(\Delta,{\bf j})
\over \sum_{j_e}{\rm Amplitude}(\Delta,{\bf j})},
\end{equation}
where
\begin{equation}\label{amplitude}
{\rm Amplitude}(\Delta,{\bf j})=\prod_e {\rm dim}(j_e) 
\prod_t (6j)
\end{equation}
is the amplitude of Ponzano-Regge model.
This shows that ${\rm Vol}(\Delta,{\bf j})$
indeed has the interpretation of the volume of a labelled
triangulation. Note that the volume turns out to be purely
imaginary. This has to do with the fact that in (\ref{zbfl3d})
one sums over configurations of $E$ of both positive
and negative volume.
A more detailed explanation of this is given in \cite{Volume}.

The volume (\ref{vol-1}) has three types of contributions: (i)
from vertices; (ii) from edges; (iii) from tetrahedra. It is not
hard to calculate the first two types of them. One finds
that each vertex contributes exactly $1/12$, and each edge contributes
$j_e(j_e+1)/6$, where $j_e$ is the spin that labels the edge $e$. 
It is much more complicated to find the
tetrahedron contribution to the volume, that is, the 
derivative of $\ln((6j)_q)$ with respect to $\Lambda$.
The result is described in the Appendix D. Using the
notations of the Appendix D, the final result for
the expectation value of the spacetime volume in BF
theory can be written as follows:

$
i{\rm Vol}(\Delta,{\bf j})= \sum_v {1\over 12} +
\sum_e {j_e(j_e+1)\over 6} 
+ \sum_t {1\over 16} {1\over \pic{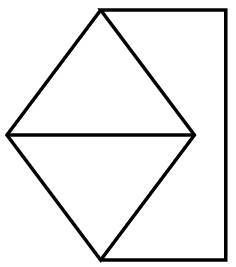}{10}{-1}{-1}{0}}\times
$ 
\begin{equation}\label{vol-2} \left( 
{1\over 24} \sum_{e\neq e'\neq e"} \left\langle \pic{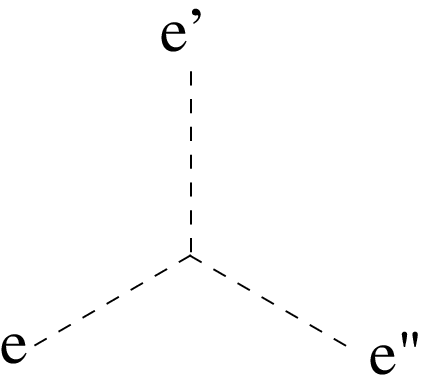}{60}{-30}{2}{2} | 
\pic{tet}{60}{-30}{2}{2} \right\rangle - {1\over 4} \sum_e \left\langle
\pic{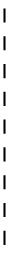}{60}{-30}{2}{2} | \pic{tet}{60}{-30}{2}{2} \right\rangle
\right)_0.
\end{equation}
Here $\pic{tet}{10}{-1}{-1}{0}$ stands for the 
normalized classical $(6j)$-symbol.

It is interesting to note that not only tetrahedra $t$ of $\Delta$
contribute to the volume, but also the edges $e$ and vertices $v$.
The contribution from the vertices is somewhat trivial -- it is constant
for each vertex. Nevertheless, when thinking about the 
triangulated manifold $\cal M$ one is forced to assign the
spacetime volume to every vertex. The contribution from edges depends on 
the spins labelling the edges. Again, this implies that each
edge of the triangulation $\Delta$ carries an intrinsic volume
that depends on its spin. The contribution from tetrahedra is more
complicated. It is given by a function that depends on the spins
labelling the edges of each tetrahedron. It is interesting that 
this picture of the spacetime volume being split into
contributions from vertices, edges and tetrahedra can be understood 
in terms of Heegard splitting of $\cal M$. Recall that Heegard
splitting of a three-dimensional manifold $\cal M$ decomposes
$\cal M$ into three dimensional manifolds with boundaries. Then
the original manifold can be obtained by gluing these manifolds
along the boundaries. For the case of a triangulated manifold 
$\cal M$, as we have now, the Heegard splitting 
proceeds as follows. First, one constructs balls centered
at the vertices of $\Delta$. Then one connects these balls
with cylinders, whose axes of cylindrical symmetry coincide
with the edges of $\Delta$. Removing from $\cal M$ the obtained
balls and cylinders, one obtains a three-dimensional manifold with
a complicated boundary. One has to further cut this manifold
along the faces of $\Delta$. One obtains three types of ``building blocks''
that are needed to reconstruct the original manifold:
(i) balls; (ii) cylinders; (iii) spheres with four discs
removed. Each of this manifolds carries a part of the
original volume of $\cal M$. Our result (\ref{vol-2})
provides one with exactly the same picture: the volume
of $\cal M$ is concentrated in vertices (balls of the
Heegard splitting), edges (cylinders), and tetrahedra
(4-holed spheres).

\subsection{Crane-Yetter model}

Crane, Kauffman and Yetter \cite{CKY} studied a state sum model 
that is very similar to the Ooguri model \cite{Ooguri}, however,
instead of the gauge group $\SU(2)$ the quantum 
group $(\SU(2))_q$ is used. The state sum invariant they
proposed is given by (see, for example, \cite{Roberts}):
\begin{equation}\label{cy}
{\rm CY}(q,\Delta) = \eta^{-2V+2E}\sum_{j_f} 
\sum_{j_t} \prod_f {\rm dim}_q(j_f) 
\prod_t {\rm dim}_q(j_t) \prod_h (15j)_q.
\end{equation}
Here ${\rm dim}_q(j)$ is the quantum dimension (\ref{qdim}),
$\eta$ is defined by (\ref{eta}),
$(15j)_q$ stands for a (normalized) quantum (15j)-symbol  (see the 
Appendix \ref{appHa}) associate with any 4-simplex.  This quantum 
(15j)-symbol is the Reshetikhin Turaev evaluation of the graph 
$\Gamma_h$ described in subsection (\ref{sec:4d}). The symbols
$V$ and $E$ stand for the 
number of vertices and edges in $\Delta$ correspondingly.  This state 
sum model is independent of a triangulation of the 4-manifold used to 
compute (\ref{cy}).  Thus, it gives an invariant associated with the 
manifold $\M$: ${\rm CY}(q,\Delta) = {\rm CY}(q,\cal M)$.

The quantity ${\rm CY}(q,\M)$ is conjectured to give the
transition amplitude for  
$BF$ theory with cosmological constant, the cosmological constant 
$\Lambda $ being related to $q$ by 
\begin{equation}\label{q-lambda}
q = \exp{i\Lambda}.
\end{equation} 
Thus, unlike in the case of 3-dimensions, we have now $\hbar = \Lambda$.
This relation can be 
established as follows.  It has been shown (see {\it e.g.} 
\cite{Roberts}) that if $\M$ is a manifold with boundary then ${\rm 
CY}(q,\M)$ is proportional to the Chern-Simons transition amplitude 
${\rm CS}(k, \partial M)$ introduced in the previous subsection:
\begin{equation}\label{d1}
{\rm CY}(q,\M) \propto {\rm CS}(k, \partial\M).
\end{equation}
The deformation parameter $q$ of the Crane-Yetter model
is related to the level of Chern-Simons theory as in
(\ref{q}).
Consider now the transition amplitude for $BF$ theory with 
cosmological term:
\begin{equation}\label{zbfl4d}
Z^{\rm BF}_4(\Lambda,{\cal M}) = 
\int {\cal D}A {\cal D}E \exp\left(
i \int_{\cal M} {\rm Tr}(E\wedge F) - {\Lambda\over 2} {\rm Tr}(E\wedge E)
\right).
\end{equation} 
Let us integrate over the field $E$ in this path integral. One gets
\begin{equation}
Z^{\rm BF}_4(\Lambda,{\cal M}) \propto
\int {\cal D}A \exp\left({i\over 2\Lambda} 
\int_{\cal M} {\rm Tr}(F\wedge F)\right).
\end{equation}
Now, using the fact that 
\begin{equation}\label{F2}
{\rm Tr}(F\wedge F) = d (A\wedge dA + {2\over 3} A\wedge A\wedge A)
\end{equation}
we get
\begin{equation}
Z^{\rm BF}_4(\Lambda,{\cal M}) 
\propto \int {\cal D}A \exp\left({i\over 2 \Lambda}
S_{\rm CS}(A,\partial\M)\right).
\end{equation}
This, together with (\ref{d1}), then implies that 
\begin{equation}
{\rm CY}(q,\M) \propto Z^{\rm BF}_4(\Lambda,{\cal M}),
\end{equation}
where $q$ is related to $\Lambda$ as in (\ref{q-lambda}).

To compare the spin foam model we obtain below with the
Crane-Yetter model we will need to know the first
order term in the decomposition of $Z^{\rm BF}_4(\Lambda,\M)$
in the power series in $\Lambda$. To find it, we
have to take into account the fact that, similarly
to the case of 3D BF theory, the transition amplitude
is only proportional to the Crane-Yetter amplitude
(\ref{cy}), with the proportionality coefficient
being a function of $\hbar$. This proportionality
coefficient is given by a power of $\hbar$. As one
can check, the ``zeroth'' order term in $\hbar$ of (\ref{cy}) is
equal to 
\begin{equation}
\left({\hbar^3\over 4\pi}  \right)^{-V+E} O(\Delta),
\end{equation}
where $O(\Delta)$ is the Ooguri state sum invariant (\ref{o}).
Thus, the first order term in $\Lambda$ of $Z^{\rm BF}_4(\Lambda,\M)$
is given by
\begin{equation}\label{cy1}
\left({\partial \over\partial \Lambda} 
Z^{\rm BF}_4(\Lambda,\M)\right)_{\Lambda=0} = 
\left({\partial \over\partial \Lambda} 
\left({{\rm CY}(q,{\cal M})\over (\hbar^3/4\pi)^{-V+E}}\right)
\right)_{\Lambda=0}.
\end{equation}
However, unlike the case of 3D BF theory with cosmological 
term, in this case the only contribution to this expression
comes from the quantum (15j)-symbol.
All other ``building blocks'' that one encounters in (\ref{cy})
contain only the $\hbar^2=\Lambda^2$ terms, and, thus, do not
contribute to (\ref{cy1}). The first derivative of the quantum
(15j)-symbol with respect to $\hbar=\Lambda$ can be
found by methods analogous to the ones used in the
Appendix D to calculate the derivative of (6j)-symbol.
However, in the case of (15j)-symbol, the calculation
is much simpler due to the fact that only crossings
contribute to the first order in $\hbar$. The (15j)-symbol
used in (\ref{cy}) contains only one crossing (see, for
example, \cite{Roberts}). Using the expression (\ref{r-matrix}) for
the R-matrix as a formal power series in $\hbar$ given
in the Appendix D, one can easily check that the
first $\hbar$-order term of the quantum (15j)-symbol
is given by:
\begin{equation}\label{cy3}
{i\hbar\over 2} \pic{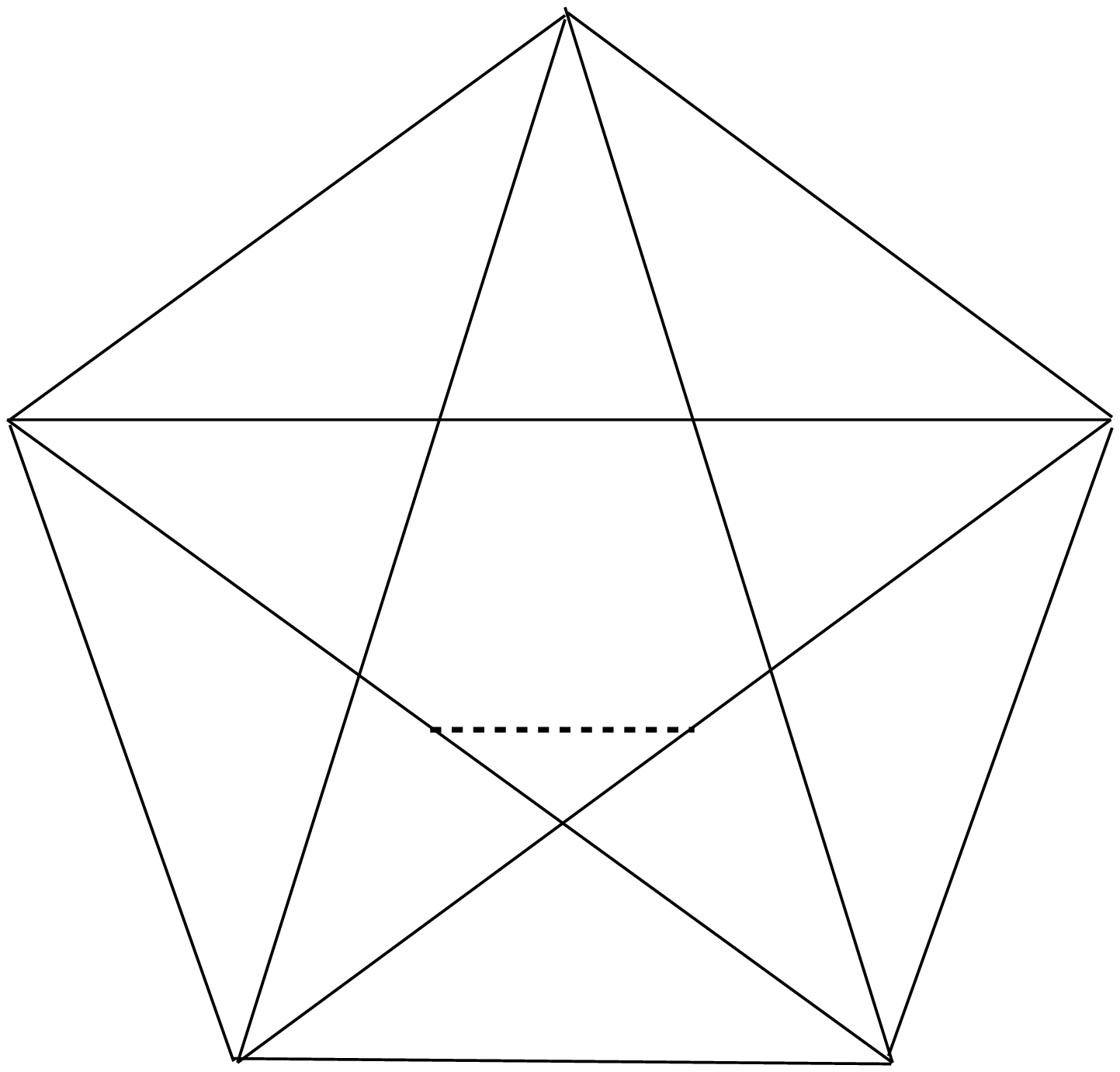}{60}{-30}{2}{2} + \cdots,
\end{equation}
where the dots stands for terms containing graspings between two edges 
sharing one vertex and correspond to different possible choices of the 
framing of the 15j-symbol.  We will not keep track of these terms at this
stage. Their relevance will be discussed below. This is, however, not quite the expression we 
want because it is not symmetrical.  We will symmetrize it by putting 
the grasping at all pairs of lines of the graph in (\ref{cy3}) that do 
not share a vertex.  However, if one does that, than the quantity one 
obtains for each given 4-simplex $h$ is not equal to (\ref{cy3}).  
Only if one sums over all 4-simplices of the triangulation one obtains 
the quantity that is equal to the sum of (\ref{cy3}) over $h$.  This 
can be proved by using certain simple relations that hold for 
graspings.  One such relation is the analogs of ``closure'' relations 
(\ref{closure}).  It graphical representation is given by:
\begin{equation}
\pic{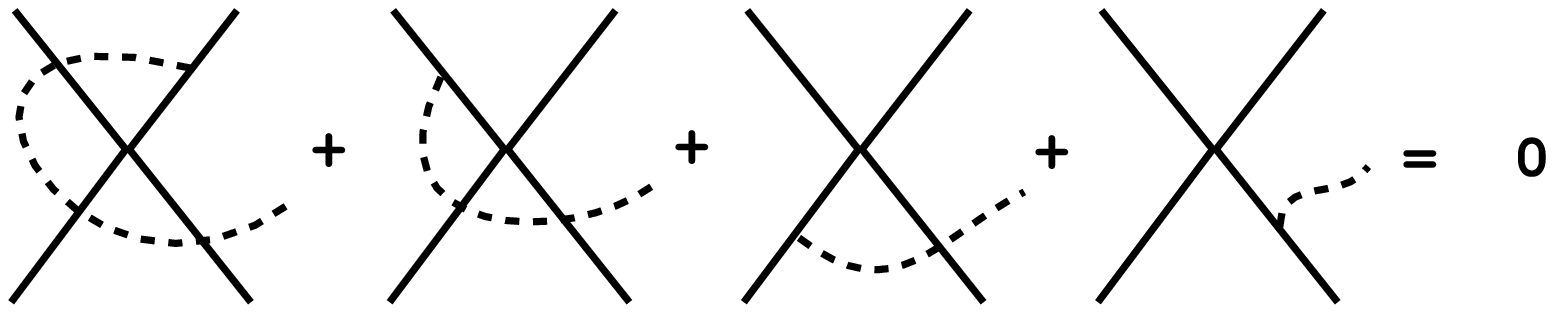}{60}{-30}{2}{2}
\end{equation}
Such ``closure'' relation holds for every vertex of the 
graph in (\ref{cy3}). Another relation that one has to use 
involves two different pentagon graphs. As one can convince oneself,
the vertices of the pentagon graph in (\ref{cy3}) are
in one-to-one correspondence with the tetrahedra of
$\Delta$. Thus, every tetrahedron is shared exactly by
two 4-simplices, and the following relations holds:
\begin{equation}
\label{pass}
\pic{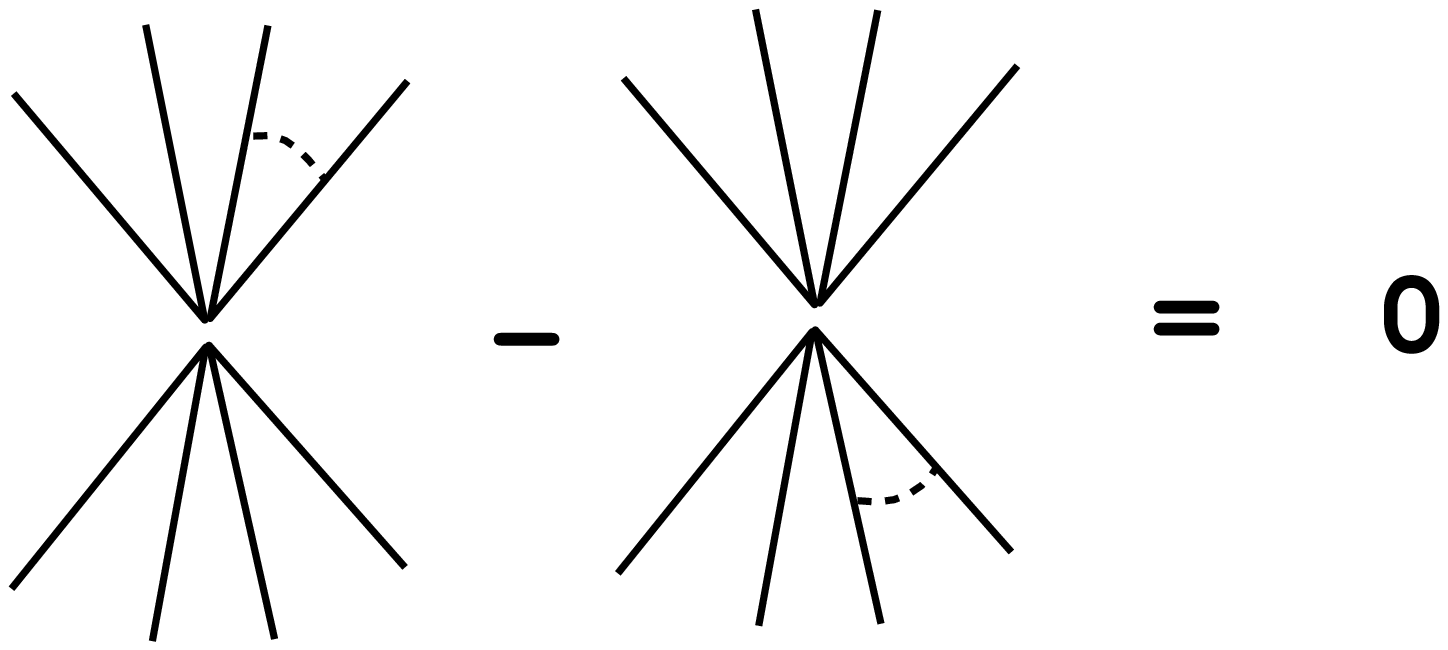}{60}{-30}{2}{2}
\end{equation}
This allows one to cancel certain types of graspings from
one 4-simplex with similar types of graspings from the
neighboring 4-simplex. 
Certain cancellations occur for each tetrahedron of the 
triangulation. These relations allow one to obtain the 
following symmetric expression:
\begin{eqnarray}\label{cy2}\nonumber
&&\left({\partial \over\partial \Lambda} 
Z^{\rm BF}_4(\Lambda,\M)\right)_{\Lambda=0} = \\ && {i\over 2}
\sum_h {1\over 30}  \left( 
\sum_{e,e'}  {\rm sign}(e,e') \left\langle
\pic{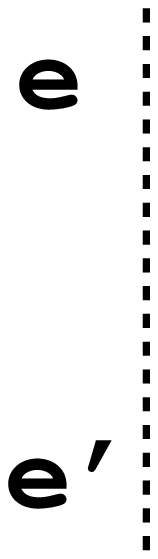}{60}{-30}{2}{2} | \pic{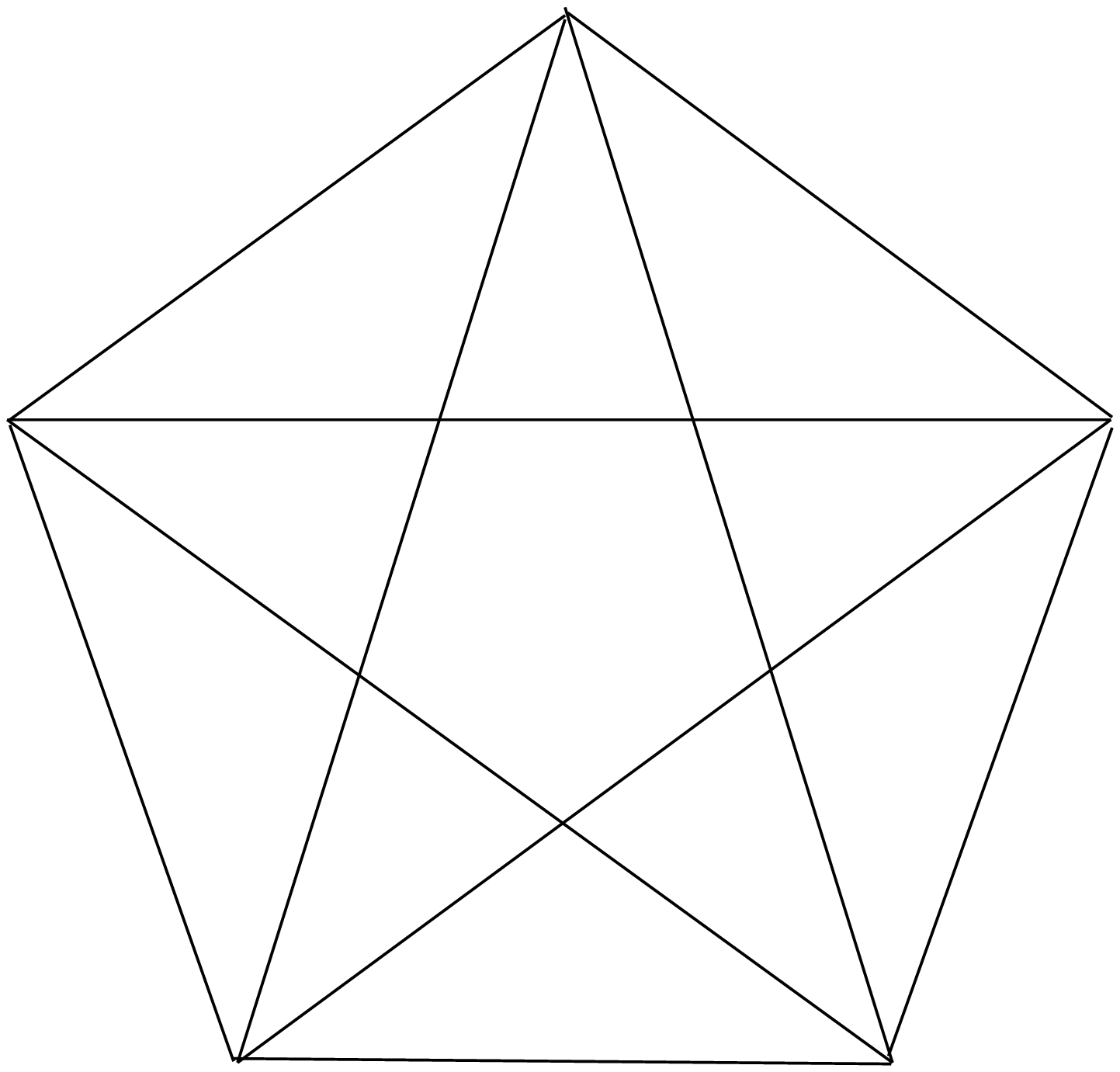}{60}{-30}{2}{2} 
\right\rangle
\right)_0 + \cdots,
\end{eqnarray}
where the sum is taken over different pairs $e,e'$ of edges
that do not share a vertex. There are 30 terms
in this sum, which explains the factor of $1/30$ in (\ref{cy2}). 
The quantity ${\rm sign}(e,e')$ in (\ref{cy2}) is plus or
minus one depending on the orientation of the two edges $e,e'$.
A consistent choice of orientations comes from the
geometrical 4-simplex discussed in the Appendix E. With this
choice, ${\rm sign}(e,e')={\rm sign}(f,f')$, where we use the fact 
that every edge of the pentagon graph in (\ref{cy2}) is
in one-to-one correspondence with a face of the corresponding
4-simplex, and ${\rm sign}(f,f')$ is defined by the
equation (\ref{bi-volume1}).
The dots in (\ref{cy2}) correspond to some symmetric expression 
that contains only terms with 
graspings between two edges sharing one vertex. 

\subsection{Reisenberger model}

Reisenberger \cite{Mike97} has proposed a state sum model corresponding to the 
self-dual Plebanski action (\ref{sdgraction}). The main idea of his 
construction is to modify the $\SU(2)$ Crane-Yetter model in
such a way that the constraints (\ref{cons}) are implemented. The model can 
be described as follows. As in the previous subsection, let $\M$ 
be a triangulated 4-manifold. Let us associate a group element 
$g_w$ to each wedge $w$ of the dual triangulation (see Section 
\ref{sec:gf}).  One can think of $g_w$ as the holonomy of connection 
around the boundary of the wedge $w$ with the basepoint being the center 
of the 4-simplex $h$ to which the wedge belongs.  Let $D_w^i$ be a 
differential ${\rm su}(2)$ operator acting on functions of $g_w$ as 
the sum of left and right invariant vector fields.  Let us denote by 
$\Omega^{ij}$ the following operator acting in $(\SU(2))^{\otimes 
10}$:
\begin{equation}
\Omega^{ij} = \sum_{w,w'} \epsilon(w,w') D_w^i 
D_{w'}^j -\frac{1}{3} \delta^{ij} 
\sum_{w,w'} \epsilon(w,w') D_w^k {D_{w'}}_k,
\end{equation} 
where $\epsilon(w,w')$ is the sign of the four volume span by 
the wedges $w,w'$ ($\epsilon(w,w')=0$ if the two wedges 
don't span a four-volume).
We choose a coloring of the wedges by spins $j_w$, denote by $R_{j}(g)$ 
the representation of $g$ as an endomorphism of $V_{j}$, and define 
for each 4-simplex $h$ the following function on $(\SU(2))^{\otimes 10}$:
\begin{equation}
\phi_{h,\vec{\jmath}}(g_w)=
tr_{\otimes_{w \in h} V_{j_w} }\left( {1\over\sqrt{2\pi x}} e^{-\frac{1}{2x^2} 
\Omega^{ij}\Omega_{ij}} \otimes_{w \in h V_{j_w}} (2j_w +1) 
R_{j_w}(g_w) \right).
\end{equation}  
The state sum model proposed by Reisenberger is given by
\begin{equation}
\lim_{x \rightarrow 0} \sum_{j_w} \int \prod_w dg_w \prod_{h} 
\phi_{h,\vec{\jmath}}(g_w)
\end{equation}
The limit is a possible way of selecting only the states belonging 
to the kernel of $\Omega^{ij}$.

\section{Applications of the generating functional}
\label{sec:appl}

In this section we use the generating functional computed in Sec.\ref{sec:gf} 
to construct systematically state sum models of the theories 
described in Sec.\ref{sec:th}.  We then compare the results of this 
systematic approach with the known results described in Sec.\ref{sec:sf}.

\subsection{BF theory: any dimension}
\label{sec:bf}

The transition amplitude of the theory is simply the value of the 
generating function $Z$ at $J=0$. As one can easily see, this 
gives exactly the Ponzano-Regge model (\ref{pr}) in three dimensions,
and the Ooguri model (\ref{o}) in four dimensions. The
corresponding two dimensional model is somewhat trivial: the transition 
amplitude is given by (\ref{z2d}) at $J=0$, which gives
\begin{equation}\label{zbf2d}
\sum_j ({\rm dim}_j)^{\chi(\cal M)},
\end{equation}
where $(\cal M)$ is the Euler characteristics of $\cal M$.
One can recognize in this expression the volume of the space of
flat connections on $\cal M$ modulo gauge transformations expressed in 
term of Riemann zeta function (see, for example, \cite{Blau}).

\subsection{2D Yang-Mills}
\label{sec:ym}

Our general strategy is to calculate the `interaction' term $\Phi(E)$
of the action (\ref{type}) on the configuration of $E$ that is
distributional along the wedges of the dual complex
\begin{equation}\label{anz}
E = \sum_w E_w,
\end{equation}
express the result as a polynomial function $\Phi(X)$ of the variables 
$X_w$, introduced in the previous section, and then look for the
vacuum-vacuum transition amplitude of the theory as given by
\begin{equation}
\left(e^{i\Phi(-i\delta/\delta J)}\,Z[J,\Delta]\right)_{J=0}.
\end{equation}

In the case of 2d Yang-Mills theory, the `interaction' term 
$i\Phi$ is given by (see (\ref{ympart}))
\begin{equation}
{e^2\over 2} \int_{\cal M} d\mu {\rm Tr}(E^2),
\end{equation}
Calculating this on the configuration of $E$ given by
(\ref{anz}), with each $E_w$ being constant along the 
wedge $w$, we get
\begin{equation}
{e^2\over 2} \sum_w \rho_w {\rm Tr}(X_w^2) =
-e^2 \sum_w \rho_w X_w^i X_w^i,
\end{equation}
where $\rho_w$ is the area of the wedge $w$, as
measured with respect to $d\mu$, and
we have introduced the $\SO(3)$ indices (\ref{components}).
The partition function of the theory is then given by
\begin{equation}
\left(\exp\left( - e^2 \sum_w \rho_w \left( 
{\delta\over\delta iJ_w^i}{\delta\over\delta iJ_w^i}\right)\right) 
\,Z[J,\Delta]\right)_{J=0}.
\end{equation}

To compare this with the partition function
of the Migdal-Witten model, it is more convenient to take
the generating functional $Z[J]$ in its general form
(\ref{k1}). Then, using
\begin{equation}\label{ymq3}
e^{\alpha \left( {\delta\over\delta iJ_w^i}{\delta\over\delta iJ_w^i}\right) }
\cdot P(J) R_{(j)}(e^J)|_{J=0} = e^{\alpha (j+1/2)^2 } = e^{\alpha 
(c(j)/2+1/4)}
\end{equation}
where $R_{(j)}(e^J)$ is the group element $e^J$ taken in the $j$
representation, and $c(j)$ is the quadratic casimir of this 
representation. We get for the partition function
\begin{equation}
\prod_\epsilon \int dg_\epsilon \sum_{j_\sigma} 
\prod_\sigma {\rm dim}_{j_\sigma} 
\chi_{j_\sigma}(g_{\epsilon_1}\cdots g_{\epsilon_n})
\exp\left(\sum_{w\in\sigma} - e^2 \rho_w (c(j_\sigma)/2+1/4) \right).
\end{equation}
This can be seen to be equal to 
\begin{equation}
{\rm YM}(\rho e^2,{\cal M}) e^{-e^2\rho/4}
\end{equation}
by noting that 
$$\sum_{w\in\sigma} \rho_w = \rho_\sigma, \qquad {\rm and} \qquad
\sum_\sigma \rho_\sigma = \rho.$$
Thus, our approach gives the correct expression
(\ref{ym}) for the partition function of the Yang-Mills theory,
apart from the factor $\exp(-e^2\rho/4)$. However, the later
is what is called a ``standard renormalization'' of the 
partition function. That is, the result for the partition function of 
2D Yang-Mills may depend on a regularization procedure that
was used to calculate it, the two different schemes giving results
that may differ, in particular, by factors of $\exp(-\alpha e^2\rho)$,
where $\alpha$ is some coefficient. See, for instance, \cite{Witten}
for a more detailed discussion of the standard renormalizations.
Thus, the partition function obtained by our method differs
from that of the Migdal-Witten model just by a standard
renormalization factor.

\subsection{3D BF theory with cosmological term}
\label{sec:bf3d}

As one can see from (\ref{bf3daction}), the `interaction' term
$i\Phi(E)$ for this theory is given by 
\begin{equation}\label{bfl1}
i\Phi(E)=-{i\Lambda\over 12}\int_{\cal M} {\rm Tr}(E\wedge E\wedge E).
\end{equation}
As in the case of Yang-Mills theory in two dimensions, one
has to find a polynomial function $\Phi(X)$ of variables
$X_w$ by evaluating (\ref{bfl1}) on the distributional field $E$
given by (\ref{distr}). Unfortunately, in the case of three dimensions the
result is not  well-defined as in the case of 
2D Yang-Mills theory. Indeed, the $E$ field is concentrated
along the wedges (see Fig. 7), and non-trivial contributions
to the integral (\ref{bfl1}) come from the points where
wedges intersect. Thus, the contributions to (\ref{bfl1})
come from the integrals
\begin{equation}\label{inters}
\int_{\cal M} {\rm Tr}(E_{w_1}\wedge E_{w_2}\wedge E_{w_3}),
\end{equation}
where $w_1,w_2,w_3$ are three wedges that intersect. The
wedges can intersect at points, as, for example,
they do at the center of each tetrahedron, or they can have 
more general intersections, as, for example, an intersection
of three wedges sharing a dual edge. Or, instead, one can have
in (\ref{inters}) $w_1=w_2$. Whatever the type of an intersection is, 
the result of (\ref{inters}) is ill-defined because of the 
distributional nature of $E$ field. In this paper we show
how the ambiguity in (\ref{inters}) can be resolved for
the simplest, and most important type of intersections:
intersections of three different wedges at the centers of
tetrahedra. As we shall see, for intersections of this type
there is a simple way to resolve the ambiguity in (\ref{inters})
using geometrical considerations. As we show, this type
of intersection is the most important one, for it is 
responsible for, in certain sense, the most intricate contribution to the
transition amplitude of the theory. Thus, in this subsection,
we restrict our consideration only to this special
type of intersection, calculate the corresponding interaction
term $\Phi(X)$, find the corresponding spin foam model,
and compare it with the Turaev-Viro model. In the first order
in the decomposition of Turaev-Viro amplitude in power of $\Lambda$, 
we will be able to identify a term analogous to the term
we obtain in our model and compare them. We will find a
very delicate matching between the two, including the
numerical coefficients. The importance of other types
of intersections, not treated here, will be emphasized later.

Thus, the intersections we consider are the ones for which
three different wedges intersect at the center of a tetrahedron.
Given a tetrahedron $t$, there are 20 (without counting the
permutations) different triples of wedges $w_1\not= w_2\not= w_3$.
Four of these triples do not span a three-volume; thus,
it is natural at first to take the integral (\ref{inters}) for such triples 
to be  zero.  These are exactly the triples of wedges that share a line -- 
a part of the dual edge.  Thus, there are only 16 different triples of 
wedges (without counting the permutations) that contribute to 
(\ref{inters}) for a given tetrahedron.  Recall now that each $E_w$ is 
given by (\ref{distr-2}).  Thus, for any given triple of wedges 
$w_1\not= w_2\not= w_3$, the integral (\ref{inters}) is proportional 
to
\begin{equation}\label{bfl2}
i\Lambda {\rm Tr}(X_{w_1}X_{w_2}X_{w_3}),
\end{equation}
but the proportionality coefficients are not fixed, because
of the indeterminacy in the value of the integral 
\begin{equation}\label{bfl3}
\int \delta(u_1)\delta(u_2)\delta(u_3) du_1\wedge du_2\wedge du_3.
\end{equation}
Of course if the wedges were infinite planes without boundaries this 
integral is a well-defined quantity: it is just the 
intersection number and as such should be 
$\pm 1$ depending on the orientations of the wedges. 
Because of the existence of boundary we expect this integral to be a 
number smaller than $1$.  We fix the proportionality coefficient by 
requiring that
\begin{equation}\label{bfl4}
{1\over 12}\int_t {\rm Tr}(E\wedge E\wedge E),
\end{equation}
where the integral is taken over a tetrahedron $t$, 
is equal to the geometrical volume of $t$. For any triple of
wedges $w_1\not= w_2\not= w_3$, the integral (\ref{bfl3})
is equal to $\alpha\cdot{\rm sign}(w_1,w_2,w_3)$, where
$\alpha$ is a coefficient that is independent on a triple,
and ${\rm sign}(w_1,w_2,w_3)$ is plus or minus one depending
on the orientation of the form $du_1\wedge du_2\wedge du_3$.
Thus, (\ref{bfl4}) is equal to 
\begin{equation}
{1\over 12}\sum_{w_1\not= w_2\not= w_3\in t}
\alpha\cdot {\rm sign}(w_1,w_2,w_3) {\rm Tr}(X_{w_1}X_{w_2}X_{w_3}).
\end{equation}
Here the sum is taken over triples of different wedges in $t$,
taking into account all different permutations of $w_1,w_2,w_3$.
Taking into account the fact that the number of permutations
of $w_1,w_2,w_3$ is 6, we get for (\ref{bfl4}):
\begin{equation}\label{bfl5}
{1\over 2}\sum_{w_1 < w_2 < w_3}
\alpha {\rm Tr}(X_{w_1}X_{w_2}X_{w_3}),
\end{equation}
where the notation $w_1 < w_2 < w_3$ means that the 
sum is taken over 16 different triples $w_1\not= w_2\not= w_3$
such that ${\rm sign}(w_1,w_2,w_3)=1$. 

To relate (\ref{bfl5}) to the volume of tetrahedron $t$, we
recall the geometrical interpretation of variables $X_w$.
They were introduced in the previous section as the
variables that carry information about the length of
edges of the triangulation. At this stage it is
more convenient to introduce  $\SO(3)$ indices
(\ref{components}). Thus, each $X_w$ is characterized
by $X_w^i, i=1,2,3$. Recall that wedges $w$
are in one-to-one correspondence with  edges $e$ of
the triangulation. Thus, let us view each
$X_w^i$ as the vector representing the corresponding edge
(each edge can be viewed as a vector pointing from one
vertex of $\Delta$ to another), and the norm squared  $X_w^i X_w^i$ 
of $X_w^i$ as the length squared of this vector.
Indeed, it is not hard to check that the 
interpretation of  $X_w^i X_w^i$ as the
length of the corresponding edge is consistent
with the other known facts. Let us consider the operator corresponding
to $X_w^i X_w^i$. According to our general prescription,
this quantity is represented by the operator
$(\delta/\delta iJ_w)^2$. We have already
dealt with this operator in the previous subsection,
see (\ref{ymq3}). Its eigenvalue is given 
just by the half of the Casimir (plus $1/4$). 
Thus, in the sense of eigenvalues, we can write
\begin{equation}\label{bfl6}
X_w^i X_w^i = (j+1/2)^2,
\end{equation}
where $j$ is the spin from (\ref{z3}) labelling the dual
face that contains $w$. Thus, (\ref{bfl6}) tells us
that, in the limit of large spins $j$, the norm
of $X_w$ grows as $j$. This is
to be compared with the length spectrum of the 
canonical quantum theory
\begin{equation}\label{lengthspec}
({\rm length}) = \sqrt{j(j+1)}.
\end{equation}
This also grows as $j$ for large spins. The expression (\ref{lengthspec})
can be easily derived in the context of
canonical (loop) quantum gravity in three dimensions
(note that we use units in which $8\pi G=1$). 
Another motivation for interpreting the spin $j$ as the length 
of an edge (for large $j$) is that it is exactly this interpretation
that must be used to reproduce correctly
the Regge calculus version of Einstein-Hilbert
action in the Ponzano-Regge model of quantum
gravity \cite{PR}. Thus, we learn that the 
interpretation of the norm of each vector $X_w$
must be that of the length of the corresponding
edge of $\Delta$. Having this fixed we can relate (\ref{bfl5}) 
to the volume of tetrahedron $t$. We have
\begin{equation}\label{bfl7}
{\rm Tr}(X_{w_1} X_{w_2} X_{w_3}) = 
2 \epsilon_{ijk} X_{w_1}^i X_{w_2}^j X_{w_3}^k,
\end{equation}
where we have introduced the $\SO(3)$ indices, see (\ref{components}). 
Each $X_w^i$ has the interpretation of the vector 
corresponding to one of the edges of $t$.
Thus, (\ref{bfl7}) is equal to $12V$, where $V$
is the volume of $t$. In (\ref{bfl5})
we have 16 such terms. Thus, it is equal to 
$${1\over 2} \alpha\cdot 16\cdot 12V.$$
The requirement that (\ref{bfl4}) is
equal to the volume of $t$ fixes the 
parameter $\alpha$ to be $1/(16\cdot 6)$. Thus, finally,
we obtain the interaction term $i\Phi(X)$ to be
\begin{equation}\label{bfl8}
-{i\Lambda\over 6}{1\over 16}\sum_{w_1 < w_2 < w_3}
\epsilon_{ijk} X_{w_1}^i X_{w_2}^j X_{w_3}^k,
\end{equation}
where the sum is taken over 16 terms. To obtain the
transition amplitude of the theory we have to
replace each $X_w^i$ by the
operator $\delta/\delta iJ_w^i$ and act by
the exponential of (\ref{bfl8}) on the
generating functional. Thus, the
first order term in $\Lambda$ 
in the decomposition of the transition amplitude is given by:
\begin{equation}\label{bfl9}
\left(\left(-{i\Lambda\over 24}{1\over 16}\sum_{w_1 < w_2 < w_3}
4 \epsilon_{ijk} {\delta\over\delta iJ_{w_1}^i} 
{\delta\over\delta iJ_{w_2}^j} 
{\delta\over\delta iJ_{w_3}^k}\right)
Z_3(J,\Delta)\right)_{J=0}. 
\end{equation}
This is to be compared with (\ref{vol-2}). We will now show 
that the most complicated term in that expression --
the term that involves trivalent graspings -- exactly
matches our result (\ref{bfl9}), including the numerical
coefficient and the sign. To see this we just have to relate
the trivalent grasping in (\ref{vol-2}) to the cubic
operator in (\ref{bfl9}). As  explained in the
Appendix D, a single grasping in (\ref{vol-2})
acts by inserting $(\sigma^i/\sqrt{2})$, where
$\sigma^i$ are the Pauli matrices. The operator $(\delta/\delta iJ^i)$,
when applied only one time, acts by inserting just $\sigma^i$.
Also, taking into account the definition of 
the trivalent grasping in (\ref{vol-2}), one can show
that 
\begin{equation}
\pic{grasp}{60}{-30}{2}{2} = 
4 \epsilon_{ijk} {\delta\over\delta iJ_{e}^i} 
{\delta\over\delta iJ_{e'}^j} 
{\delta\over\delta iJ_{e''}^k}
\end{equation}
So  our approach {\it does} account for the most intricate part of 
the 3D transition amplitude in the first $\Lambda$-order, reproducing 
correctly even the numerical coefficients.  Note, however, that we do not 
get all of the terms appearing in (\ref{vol-2}).
The discrepancy we  find between the model we obtain
and the usual state sum model can be
understood by comparing the two different expressions 
for the spacetime volume: the one obtained within our approach
and the one obtained within the Turaev-Viro model, see (\ref{vol-2}).
The Turaev-Viro model tells us that we must associate the
spacetime volume not only to tetrahedra of the triangulation,
but also to edges and vertices. Within our approach, only
the tetrahedron part of the spacetime volume is accounted
for. The reason for this was that, when evaluating 
the cosmological term of the action on the distributional 
$E$ field, we took only the terms which came from 
intersections of different wedges at the centers of tetrahedra.
As we discussed, even
these terms are ill-defined because of
the distributional nature of the $E$ field. However, the
corresponding ambiguities can be successfully 
resolved for these types of intersections by using geometrical
considerations, and, after the ambiguities are eliminated,
the result exactly matches the 
analogous terms of the usual state sum models, including
the matching of the numerical factors. The geometrical considerations
we used were exactly the ones that relate the terms we
considered to the volume of a geometrical 3-simplex.
Thus, it is not very surprising that only the ``part'' of the Turaev-Viro
model that accounts for the volume of 3-simplices was 
reproduced correctly: as the geometrical considerations we
used above tell us, we considered
only the terms that are relevant for the volume of
individual 3-simplices. However, there are
other types of terms that we did not consider and that
may be crucial to reproduce the corresponding state sum model
correctly. Let us consider, for example, the terms of the
type:
$$\int_t {\rm Tr}(E_{w_1}\wedge E_{w_1}\wedge E_{w_2}),$$
for some $w_1\not=w_2$. The result of such an integral
is ill-defined. Indeed, there is an indeterminacy 
of the type $0\cdot\infty$, where $0$ comes from 
$du_1\wedge du_1$ (see (\ref{distr-2})), and $\infty$ comes
from the square of the $\delta$-function $\delta^2(u_1)$.
These terms are proportional to
$${\rm Tr}(X_{w_1}X_{w_1}X_{w_2}).$$

Other terms that may arise are
$$\int_t {\rm Tr}(E_{w}\wedge E_{w}\wedge E_{w}),$$
for some $w$. They arise as the result of the
indeterminacy $0^2\cdot\infty^2$, where $0^2$ comes from 
$du\wedge du\wedge du$ (see (\ref{distr-2})), and $\infty^2$ comes
from the cube of the $\delta$-function $\delta^3(u)$.
These terms are proportional to
$${\rm Tr}(X_{w}X_{w}X_{w}).$$
There is no obvious reason to set these two types of 
terms to zero. In fact, as one can see, these are
exactly the types of terms that are needed to account
for the other terms appearing in the first $\Lambda$-order
in the Turaev-Viro model.
Although the structure of terms appearing this
way is clear, at the present
state of the development of our approach, no 
geometrical arguments is available to fix the
ambiguity in the coefficients in front of such terms.
However, the precise agreement of the terms for which 
such arguments do exist gives hope that, ones the above 
ambiguities are resolved, we will have an exact agreement
between the models obtained using our procedure and the usual 
state sum models.

\subsection{4D BF theory with cosmological term}
\label{sec:bf4d}

In the case of 4D BF theory with cosmological term 
the interaction $i\Phi$ is given by:
\begin{equation}
i\Phi(E)=-{i\Lambda\over 2}\int_{\cal M} {\rm tr}(E\wedge E).
\end{equation}
Our general prescription is to evaluate this interaction term
on the distribution (\ref{distr-1}) and find a polynomial function
$\Phi(X)$. Thus, one has to evaluate integrals
\begin{equation}\label{bf4-1}
\int_{\cal M} {\rm Tr}(E_w\wedge E_{w'})
\end{equation}
with $E_w$ given by (\ref{distr:4d}). The integral (\ref{bf4-1})
is non-zero only if wedges $w,w'$ intersect. However, similarly
to the case of 3D, the result when the wedges intersect is ill-defined
because of the distributional nature of $E_w$. Thus, again some
independent considerations have to be used to fix the 
ambiguity. We use a strategy similar to the one adopted in 
the case of 3D BF theory. We consider only the terms coming from
wedges $w,w'$ intersecting at the center of a 4-simplex. As we
shall see, these are the most important terms in the sense
that they are responsible for the main terms in the first order in $\Lambda$
of the Crane-Yetter model. The relevance of other types of terms will
be emphasized below. To fix the ambiguity in 
(\ref{bf4-1}) when $w,w'$ are two wedges that intersect at
the center of a 4-simplex $h$ we use geometrical considerations.

Let us consider the integral
\begin{equation}\label{bf4-2}
- {1\over 2} \int_h {\rm Tr}(E\wedge E)
\end{equation}
over the interior of a particular 4-simplex $h$. It is equal to the
sum
\begin{equation}
- {1\over 2}\sum_{w,w'\in h} \int_h {\rm Tr}(E_w\wedge E_w').
\end{equation}
Each of the integrals here is proportional to 
\begin{equation}
{\rm Tr} (X_w X_{w'}) = - 2 X_w^i X_{w'}^i,
\end{equation}
with the proportionality coefficient given by:
\begin{equation}\label{bf4-3}
-{1\over 2}\int_h \delta(u) \delta(v) \delta(u') \delta(v')
du\wedge dv\wedge du'\wedge dv'.
\end{equation}
The later is equal to $-(1/2)\alpha\cdot{\rm sign}(w,w')$,
where $\alpha$ is a numerical parameter, whose value is not
fixed due to the ambiguity referred to above, and ${\rm sign}(w,w')$
is the sign of the volume form in (\ref{bf4-3}). Thus, (\ref{bf4-2})
is equal to
\begin{equation}\label{bf4-4}
\sum_{w,w'\in h} \alpha\cdot{\rm sign}(w,w') X_w^i X_{w'}^i,
\end{equation}
where the sum is taken over wedges $w,w'$ inside $h$
that span a 4-volume. There are exactly 30 terms 
summed over in (\ref{bf4-4}).

We will fix the parameter $\alpha$ relating the quantities
$X_w^i$ to the geometrical 4-simplex in $\R^4$. First, let
us note that when $E$ in (\ref{bf4-2}) is equal to
the self-dual part of the wedge product of two copies
of the frame field:
\begin{eqnarray}\nonumber
E_{ab} = {}^+\Sigma_{ab}, \\
\Sigma_{ab}^{IJ} = \theta^I_{[a} \theta^J_{b]},
\end{eqnarray}
where $\theta^I_a$ is a frame (tetrad) field,
then (\ref{bf4-2}) is equal to 
\begin{eqnarray}\label{bf4.2}\nonumber
\int_h {1\over 2} {}^+\Sigma_{ab}^i
{1\over 2} {}^+\Sigma_{cd}^i \tilde{\epsilon}^{abcd} &=&
\int_h {1\over 2} {}^+\Sigma_{ab}^{IJ}
{1\over 2} {}^+\Sigma_{cd\,IJ} \tilde{\epsilon}^{abcd}\\ &=& 
{1\over 16} \int_h \epsilon_{IJKL} \Sigma_{ab}^{IJ}
\Sigma_{cd}^{KL} \tilde{\epsilon}^{abcd} = {4!\over 16} V_h,
\end{eqnarray}
where $V_h$ is the volume of $h$ with 
respect to the metric defined by $\theta^I_a$.
Thus, we will fix $\alpha$ in such a way that 
(\ref{bf4-4}) is equal to $(3/2)V_h$ when
$X^i_w$ can be related to the quantities
characterizing a geometrical 4-simplex.

Recall that a
geometrical 4-simplex in $\R^4$ is characterized (up
to translations) by four vectors: the four vectors pointing
from one of the vertices to the other four ( for more details
see the Appendix E). For each face
of $h$ one can also construct the so-called bivectors,
which are given by the wedge products of any two of the
edges belonging to that face. The bivectors live in the
second exterior power of $\R^4$. One can take the self-dual
part of a bivector to obtain an element of $\R^3$. 
As is shown in the Appendix E (see (\ref{biselfarea})), the norm
(obtained using the usual flat metric in $\R^3$) of the
self-dual part of each bivector is equal to the squared area
of the corresponding face. There exists also a 
simple expression (\ref{sd4volume}) for the 
volume of a 4-simplex involving only the self-dual
parts of bivectors. Recall
now that wedges $w\in h$ are in one-to-one correspondence
with faces of $h$. Then, if $X_w^i$ has the interpretation of
the self-dual part of the bivector corresponding to a face
of $h$.
 
A comparison of (\ref{sd4volume}), (\ref{bf4.2})
fixes the value of $\alpha$ to be $\alpha=1/30\cdot 4$.
Thus, the first $\Lambda$-order term of the 
transition amplitude is given by:
\begin{equation}\label{bf4-5}
 \left(\Lambda {1\over 30} \sum_{w,w'\in h} 
{1\over 4} {\rm sign}(w,w') {\delta\over\delta iJ_{w}^i} 
{\delta\over\delta iJ_{w'}^i} Z(J) \right)_{J=0}
\end{equation}
To compare this with (\ref{cy2}) we need the relation between the 
grasping there and the operator in (\ref{bf4-5}).  The corresponding 
relation is given by
\begin{equation}
\pic{grasp2dif}{60}{-30}{2}{2} = {1\over 2}
{\delta\over\delta iJ_{w}^i} 
{\delta\over\delta iJ_{w'}^i} 
\end{equation}
Using this correspondence we see that (\ref{bf4-5}) agree with the 
result (\ref{cy2}). As in the 3D case, the matching includes the
numerical coefficient and the sign of the expressions.

Let us now discuss the role of the terms denoted by dots in (\ref{cy2}). 
Let us recall that those terms are determined by a framing of the
(15j)-symbol used in the Crane-Yetter model. These terms are
given by a sum of graspings of edges of the graph $\Gamma_h$
that share a vertex. Recall that vertices of $\Gamma_h$ are
in one-to-one correspondence with the tetrahedra of h. Thus,
the general structure of these terms is
such that they can be grouped according to a 3-simplex (tetrahedron)
to which they ``belong''. Therefore, these terms can be thought of
as the contribution to the 4-volume of the BF theory coming from
individual tetrahedra of $\Delta$. Thus, these terms are in certain
sense analogous to the terms in 3D that carry 3-volume corresponding
to the edges of $\Delta$. As in 3D our approach did not reproduce
correctly the terms corresponding to edges, in 4D we did not
reproduce correctly the contribution to the 4-volume coming 
from tetrahedra. Similarly to the case of 3D, these terms may
possibly be reproduced if one takes into account the types
of intersections in (\ref{bf4-1}) other than the ones considered here.

\subsection{4D Euclidean gravity}
\label{sec:gr}

In order to apply our strategy to the self-dual Plebanski model 
let us consider the ``interaction'' term of the action (\ref{sdgraction}):
\begin{equation}\label{sdgr-1}
i\Phi(E) = -i \int_{\M} \psi_{ij} \left( 
E^{i}\wedge E^{j} - {1\over3}\delta^{ij} E^k E_k \right).
\end{equation}
Using the same procedure as in the case of the 4-dimensional BF theory with 
cosmological constant, this term, when evaluated on distributional $E$ 
field, gives 
\begin{equation}\label{sdgr-2}
-i \sum_{h} \psi_{ij}(v_h) \Omega^{ij}_h(X)
\end{equation}
where $v_h$ denotes the center of the 4-simplex and
\begin{equation}\label{sdgr-3}
\Omega^{ij}_h(X)= \sum_{w,{w'}\in h} {\rm sign}(w,{w'}) \left(
X_w^i X_{w'}^j - {1\over 3}\delta^{ij} X_w^k X_{w'\,k} \right)
\end{equation}
is a quadratic function in the wedge variables $X_w$. 
In order to write (\ref{sdgr-2})
we have absorbed some unimportant numerical constants into the Lagrange
multipliers $\psi_{ij}$.
The generating functional approach tells us that the spin foam transition
amplitude of the self dual Plebanski model is given by $\tilde{Z}(0)$ where 
\begin{equation}\label{sdgr-4}
\tilde{Z}(J) = \int \prod_h d\psi^{ij}(v_h)\,
e^{-i \sum_{h} \psi_{ij}(v_h) \Omega^{ij}_h({\delta \over i \delta J})}\cdot 
Z(J).
\end{equation}
The integration over $\psi^{ij}$ gives rise to delta functions. Thus, 
$\tilde{Z}(J)$ can be characterized as the solution of 
\begin{equation}\label{sdgr-5}
 \Omega^{ij}_h({\delta \over  i\delta J}) \tilde{Z}(J) = 0,
\end{equation} 
for all 4-simplices $h$, or, equivalently, as the solution of 
\begin{equation}
\Omega^{i_1j_1}_h({\delta \over  i\delta J}) \cdots 
\Omega^{i_nj_n}_h({\delta \over  i\delta J})\tilde{Z}(J) |_{J=0} =0.
\end{equation}
Using the fact that 
$Z(J)$ and, therefore, 
$\tilde{Z}(J)$ are gauge invariant and the property that an $su(2)$ 
symmetric traceless tensor $\Omega^{ij}$ is totally characterized up to 
gauge transformation by its square $ \Omega^{ij}\Omega_{ij}$ we see 
that the preceding equation is equivalent to
\begin{equation}
(\Omega^{ij}_h{\Omega_{ij}}_h)^n({\delta \over  i\delta J})\cdot 
\tilde{Z}(J)|_{J=0}=0.
\end{equation}
The solution of the later can be written as
\begin{equation}
\tilde{Z}(0) = \lim_{x \rightarrow 0} 
\prod_h \frac{1}{\sqrt{2\pi x}}
e^{-\frac{1}{2x^2}\Omega^{ij}_h{\Omega_{ij}}_h}({\delta \over  i\delta J})
\cdot Z(J)|_{J=0}
\end{equation}  

In this form the result is very similar to the one given by 
the Reisenberger model. 
There are however several differences. The first one is the fact that the
derivative operators appearing in our result are the commutative 
ones, while Reisenberger uses the non-commutative right and left 
invariant vector fields. The second difference is due to the presence 
of the factors of $P(J)$ in the generating functional. 
In the next section we will emphasize significance of these discrepancies.

\section{Discussion}

In this section we summarize what has been achieved, 
discuss open questions of the approach and give some
speculations as to possible directions of future research.

As we have said in the introduction, our main aim was to 
understand whether the known spin foam (state sum) models,
in particular the ones discussed in section \ref{sec:sf},
could be obtained as a result of some systematic procedure that starts 
from the corresponding classical action 
principle. We have proposed such a procedure, motivated
by certain results of the loop approach to quantum gravity.
Our main idea is two-fold. First, we proposed to include in
the path integral certain distributional field configurations.
In fact, in this paper we concentrated only on such
distributional configurations. The distributional
field was taken to be the $E$ field. As we discussed above,
this is motivated by the distributional nature of
the non-abelian ``electric'' field of the canonical approach.
We put the distributional $E$ field on a collection of two-dimensional
polygons in spacetime, for which we used the collection of
dual cells (faces) of a fixed triangulation of the
spacetime manifold. Second, to calculate the path integral,
we have employed the idea of the generating functional. This
allows one to discuss different theories from the same point of
view. As we have seen, the generating functional is the
same for any theory in a given spacetime dimension, and
transition amplitudes for different theories arise as
different formal power series in the derivatives with respect
to the current.  Thus, this is quite reminiscent of the standard quantum 
field theory in Minkowski spacetime picture, where many different 
theories can arise from the same free theory as sums over different 
types of Feynman diagrams.  As we have seen, the role of the free 
theory is played in our approach by BF theory.  Below we shall 
speculate more on the analogy between the diagrammatic expressions 
arising in our approach and Feynman diagrams.

We have seen that our approach reproduces correctly the
transition amplitudes for BF theory in any dimension and
the partition function for Yang-Mills theory in two dimensions. 
We have also compared the transition amplitudes for BF
theories with cosmological term that arise in our approach with
the ones given by the corresponding states sum models.
We did this by comparing the first order terms in
the decomposition of the corresponding amplitudes
in the power series in $\Lambda$. In the case of 
4D theory, we found agreement (apart from the terms that
have to do with the framing), 
and in the 3D case only a partial agreement was reached. 
Note, however, that 
our approach {\it does} account for the most intricate part of the 3D and 4D 
transition amplitudes in the first $\Lambda$-order, reproducing 
correctly even the numerical coefficients.
We have also seen, both  in the cases of 3D and 4D, that the missing contributions 
can be interpreted as coming from singular contributions arising when 
one regularizes the volume operator. We expect that some new 
ideas will be necessary to resolve these regularization
ambiguities. Presumably, a study of the higher 
order terms in $\Lambda$ may give some clues as to
how these ambiguities must be resolved.

Let us conclude with some speculations and 
directions for future research. An interesting open question that
arises within our approach is as follows. We have 
seen that the state sum formulation of 
the topological BF theory can be obtained by 
using distributional fields associated 
with a 2-dimensional cellular complex. Moreover, the residual 
regularized action on the 2-dimensional worldsheet is given 
by 2D BF-Yang-Mills theory.
It is important to understand why such result is valid.
One possibility would be to use the degenerate solutions found by Baez
\cite{SpinFoams} as some monopole contributions playing a major role 
in defining the measure of the field theory. 
It might be that the topological nature of the 
theory depends on whether these singular solutions dominate or not the 
partition function.

In this paper we have discussed only correlation functions in the $E$ 
fields. It is also possible to use our approach to compute  
the expectation value of  the Wilson loop functionals. 
We hope to perform such calculations in the future.

Let us now comment on the analogy between the diagrammatic 
expressions we have found and Feynman diagrams. In our approach
the role of the ``free'' theory is played by the BF theory. Thus,
the free transition amplitudes in our case are just the BF transition
amplitudes, given by certain sums of products of functions 
of spins labelling the triangulation. This is to be compared with 
the Feynman diagram technique free amplitudes which are
just lines representing the free propagation of quanta. It is 
important that the BF amplitudes are independent of the
triangulation used to compute them. They depend only on the
properties of the initial and final states (and, of course, on the 
topology of the manifold that is contained between the
initial and final hypersurfaces). The more complicated, ``interacting''
amplitudes are given in our approach by inserting various chord
diagrams into spin networks, with spin networks representing the free 
amplitudes. This is to be compared with how one takes into account
interactions in the standard QFT's: graphically this is also represented
by inserting certain types of vertices into the free diagrams. 
In our case, each grasping occurs within a particular simplex
of the triangulation. One then has to take a sum over simplices.
Similarly, in QFT's (in the coordinate representation) one puts
an interaction vertex at a particular point and then 
integrates over the position of this point. Thus, it is tempting
to draw an analogy between a simplex of a triangulation in our
case and a point in QFT's. To summarize, spin networks play
in our theory the role of the diagrams of free propagation,
and the graspings by the chord diagrams are analogous to 
the insertion of interaction vertices into the diagrams of QFT's.

Finally, let us discuss implications of the results we have obtained for
the known spin foam models of quantum gravity in 4D. The most 
popular of these are the Reisenberger model \cite{Mike97}, discussed in some 
details above, and the Barrett-Crane model \cite{BC}. The major
problem with both of them is that they are rather invented then 
derived as a result of some systematic procedure starting from
the Einstein-Hilbert action or any equivalent of it. However,
as we have seen on the example of the Reisenberger model,
and as one can check for the case of Barrett-Crane model, 
our approach to some extent reproduces the
transition amplitudes given by these models. For example,
our approach gives a clear interpretation of the so
far mysterious evaluation of the spin networks that 
satisfy the constraint equations on the flat connection
that is performed
to obtain the amplitude. In our approach this evaluation
has an obvious meaning of taking $J=0$ at the end of the
calculation of a transition amplitude. We have also seen
that the constraint equations of the known spin foam models 
have the meaning of 
differential equations that must be satisfied by
the generating functional. However, although the
main features of the both models were reproduced by
our approach, some important details are different
in the models we have obtained. Thus, our approach
prescribes to use the commuting variational derivatives
with respect to the current instead of non-commuting 
vector fields used by Reisenberger \cite{Mike97}. Also,
the presence of the function $P(J)$, which plays a 
crucial role in, for
example, 2D Yang-Mills theory, is not accounted for 
by the Reisenberger and Barrett-Crane models.

One more important lesson that one can learn 
from our results is as follows. On examples of
3D and 4D BF theories with cosmological term
we have seen that the simplest regularization
of the ``interaction'' (volume) term of the action
that we have discussed above {\it does not}
account for all terms one gets from the 
corresponding ``exact'' models. However,
the Reisenberger model, for example, uses
exactly the same way of regularizing the constraints
of the theory, for, using the terminology of
our approach, it takes into account only
the simplest types of intersections of wedges
when regularizing the constraint part of the
action. Our experience with the BF theories
tells us that this cannot be correct. Thus, it 
could be that the problems with the Reisenberger
model (for example, no interesting solutions of
his constraint equations is known) arise simply
because this model does not take into a proper
account other terms that arise in the regularization
of the constraints. Similar is true for the 
Barrett-Crane model.

The original motivation for our research program
was to find a systematic procedure that gives one 
a way to derive spin foam models of various
theories starting from the corresponding action
functionals. However, the procedure proposed in 
this paper is still far from being able to claim to have
achieved this goal. Yet, by studying the procedure
proposed, we have learned a lot about existing models
and obtained an important insight as to their deep 
internal structure. Our results also led us to the
conclusion that the existing models of quantum gravity
in 4D are too oversimplified. In spite of these rather
negative conclusions, we believe that some of the
insights we have obtained in this paper will become a part of the final,
possibly much more elaborate picture.

{\bf Acknowledgements}: We are grateful to John Barrett, John Baez, 
Matthias Blau and Jose Zapata for discussions. K. K.
was supported in part by the NSF
grant PHY95-14240, by Braddock fellowship from Penn State and by the 
Eberly research funds of Penn State, L.F was supported by the CNRS 
and a NATO grant. 

\appendix
\section{Conventions and notations}

We use the following conventions for differential forms and
integrals:
\begin{eqnarray}
A = {1\over p!} A_{a_1\ldots a_p} dx^{a_1}\wedge\ldots\wedge dx^{a_p}, \\
\int A_{a_1\ldots a_d} dx^{a_1}\wedge\ldots\wedge dx^{a_d} =
\int d^d x A_{a_1\ldots a_d} \tilde{\epsilon}^{a_1\ldots a_d},
\end{eqnarray}
where $\tilde{\epsilon}^{a_1\ldots a_d}$ is the Levi-Civita
density, taking the values plus minus one in any
coordinate system. Our convention for the curvature form
is: $F = d A +A\wedge A$.

Throughout the paper the following symbols will stand for
the following elements of the triangulation
\begin{center}
$e$ --- for an edge\\
$f$ --- for a face\\ 
$t$ --- for a tetrahedron \\
$h$ --- for a 4-simplex
\end{center}
We will also use the symbol $\epsilon$ to denote edges of the 
dual complex (dual 1-cells), and $\sigma$ to denote dual faces
(dual 2-cells). 

All traces that we use in this paper are in the fundamental
representation.

Our $\SO(3)$ index conventions:
\begin{equation}\label{components}
X = i\tau^i X^i, \qquad J = {i\over 2} \tau^i J^i.
\end{equation}

\begin{equation}\label{eta}
\eta = {(q^{1/2}-q^{-1/2})\over i\sqrt{2k}}. 
\end{equation}
The quantity ${\rm dim}_q(j)$
is the so-called quantum dimension of $j$ ${\rm dim}_q(j)=[2j+1]_q$, 
where $[n]_q$ is the quantum number
\begin{equation}\label{qdim}
[n]_q = {q^{n/2} - q^{-n/2}\over q^{1/2} - q^{-1/2}}
\end{equation}
having the property that $[k]_q=0$.

\section{Summary of facts on $\SU(2)$}
\label{app:1}

Here we give a short summary of some standard facts about the group
$\SU(2)$. One can parameterize an element $g$ of $\SU(2)$ by
vectors $Z$ from the Lie algebra. The corresponding relation
is given by the exponentiation map:
\begin{equation}\label{app-1}
g = e^Z = e^{i\psi n^i \sigma^i/2},
\end{equation}
where $n^i$ is a unit vector $n^i n_i=1$, $\psi$ is a real positive
parameter, and $\sigma^i$ are the usual Pauli matrices:
\begin{equation}
(\sigma^i \cdot \sigma^j) = i\epsilon^{ijk} \sigma^k +\delta^{ij}.
\end{equation}
Thus, $\psi n^i$ is an element of $\R^3 \sim {\rm su}(2)$,
and (\ref{app-1}) gives the exponentiation map. As one can
easily check,
\begin{equation}
g = \cos{(\psi/2)} + i n^i \sigma^i \sin{(\psi/2}).
\end{equation}
Thus, to cover the whole $\SU(2)$ just ones, the parameter
$\psi$ should takes values in the range: $\psi\in [0,4\pi]$.

The Haar measure on the group can be related to the
usual Lebesgue measure in $\R^3$ by introducing a
function $P(Z)$ on the Lie algebra:
\begin{equation}
P(Z) = \left({\sin{(\psi/2)}\over \psi/2}\right).
\end{equation}
Then $P^2(Z) dZ/32\pi^2$ gives the normalized Haar measure
on $\SU(2)$ in terms of the Lebesgue measure $dZ$.

The characters of irreducible representations are given by:
\begin{equation}\label{char}
\chi_j(e^Z) = {\sin{(\psi(j+1/2))}\over\sin{(\psi/2)}},
\end{equation}
where $j$ are half-integers (spins).

The Fourier transform on $\SU(2)$ maps any function
on the group into a function on the space dual to 
the Lie algebra. Let $f$ be a coordinate on the
space ${\rm su}(2)^*$. Then the Fourier transform is given by:
\begin{equation} \label{Ftr}
\tilde{\phi}(f):=\int {dZ\over V} P(Z) e^{-i\,f(Z)} \phi(\exp{Z}).
\end{equation}
The inverse Fourier transform is given by:
\begin{equation} \label{IFtr}
\phi(\exp{Z}) = \sum_j {\rm dim}_j 
{1\over P(Z)}\int_j d\Omega_f e^{i\,f(Z)} \tilde{\phi}(f),
\end{equation}
where the integrals are taken over the co-adjoint orbits --
spheres of radius $j+1/2$, and the measure $d\Omega_f$ on
each orbit is normalized so that
\begin{equation}
\int_j d\Omega_f = {\rm dim}_j = 2j+1.
\end{equation}
A particular case of (\ref{IFtr}) is the following 
simple formula for characters (\ref{char}):
\begin{equation}
\chi_j(e^Z) = {1\over P(Z)} \int_j d\Omega_f e^{i\,f(Z)}.
\end{equation}
This is the famous Kirillov formula \cite{Kirillov}. 

\section{Haar measure, intertwiners and spin networks}
\label{appHa}

Let us denote by $V_i$ the spin $i$ representation of $\SU(2)$ 
and by $V_i^*$ the dual representation, which in the case of $\SU(2)$
is isomorphic to $V_i$. Let us denote by $\epsilon_i$ the corresponding isomorphism, 
and by $R_i(g)$ the representation of the group element $g$ in $V_i$.

Throughout the paper we use the normalized Haar measure 
\begin{equation}
\int dg = 1.
\end{equation}

Let $K$ be an intertwiner between the trivial representation and
a representation $V$, that is $K \in Hom_G(\R, V)$. We will denote 
by $\bar{K}$ its dual ${\bar K} \in Hom_G(V,\R^*)$. Intertwiners
are the basic building blocks of the so-called {\it spin networks}.
Another usage of the intertwiners is to express the result of 
integration of a product of group elements. 
Let us denote by $K_\alpha ^{i_1,\ldots,i_n}$ an orthonormal 
basis of $ Hom_G(\R, V_{i_1}\otimes \cdots \otimes V_{i_n})$.
The intertwiners $K_\alpha ^{i_1,\ldots,i_n}$ have the
property that 
$$\bar{K_\alpha} K_\beta= \delta_{\alpha,\beta},$$ 
where the product between $\bar{K}$ and $K$ is defined by the duality 
bracket $V\otimes V^* \rightarrow \R$.  Then the integral of $n$ group 
elements is given by
\begin{equation}
\int dg \, R_{i_1}(g)\otimes \cdots \otimes R_{i_n}(g)
= \sum_\alpha  K_\alpha ^{i_1,\ldots,i_n} {\bar{K}^\alpha}_{i_1,\ldots,i_n},
\end{equation}
where $ R_{i}(g)$ is considered as an element of $V_i\otimes V_i^*$.
To integrate a product where both $g$ and $g^{-1}$ appear 
one has to use the duality relation
$ \bar{R}_i(g)=R_i(g^{-1})= \epsilon_i {R}_i(g) \epsilon_i^*$.
For instance 
\begin{equation} \label{a1}
\int dg \,(\overline{R}_i)_{mn}\,( R_j)_{m'n'} = 
{1\over {\rm dim}_j} \delta_{ij} \delta_{mm'} \delta_{nn'}.
\end{equation}
This equality can be conveniently expressed graphically
if one represents a matrix element $(R_i)_{mn}$ by a
line labelled by $i$ with the two open ends
corresponding each to one of the indices $m,n$. Let 
us symbolically denote the integration over the group
by a circle going around the lines representing the
group elements. Then (\ref{a1}) takes the following form:
\begin{equation}
\pic{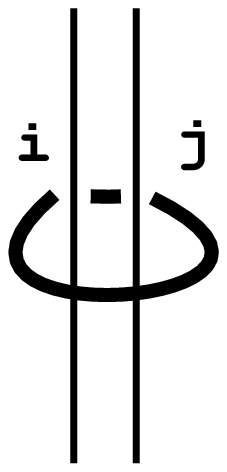}{50}{-25}{2}{2} = {1\over{\rm dim}_i} \delta_{ij} 
\raisebox{-25pt}{
\hspace{2pt}\psfig{file=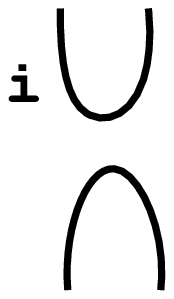,height=50pt,width=25pt,silent=}\hspace{2pt}}
\end{equation}
To find a graphical representation of the result of the
integrals involving more than two matrix elements we
introduce a trivalent vertex -- analog of the Clebsch-Gordan
symbol -- normalized in a special way. But first, let us
define the so-called {\it spin networks}.

An $\SU(2)$ spin network functional is defined by the following data:
(i) an oriented graph $\Gamma$; (ii) a map $\bf j$ from the set of edges of 
$\Gamma$ to the set of irreducible representations of $\SU(2)$; (iii)
a map $\bf i$ from the set of vertices $V$ to the set of intertwiners,
which assigns to each vertex $v$  
an intertwiner from the tensor product of representations labelling
the incoming edges to the tensor product of representations labelling the
outgoing edges. A spin network is a function on a number of
copies of $\SU(2)$, more precisely, on $(\SU(2))^E$, where
$E$ is the number of edges in $\Gamma$. To find the value of
this function one takes for each edge the matrix element 
of the group element on that edge
in the representation that labels the edge, and contracts
the matrix elements corresponding to all edges using 
intertwiners at vertices. The function obtained this 
way is gauge invariant.

In this paper we will use the normalized trivalent vertex,
defined in such a way that the $\theta$-symbol constructed
from 3 spins is always equal to unity:
\begin{equation}\label{theta-norm}
\left(\pic{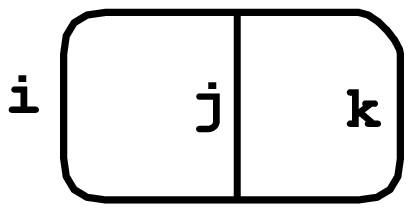}{30}{-15}{-2}{4}\right)_0=1,
\end{equation}
where the operation $(\cdot)_0$ denotes the evaluation
of the corresponding spin network
on all group elements equal to the unity in the group.

With this normalization of the trivalent vertex intertwiner,
one can show that the following relation holds:
\begin{equation}
\pic{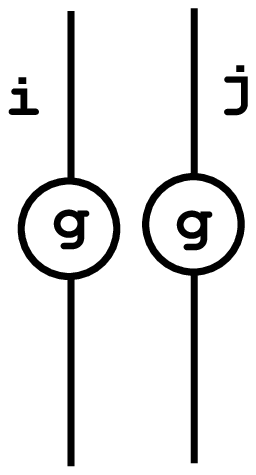}{50}{-25}{2}{2} =
\sum_k {\rm dim}_k \pic{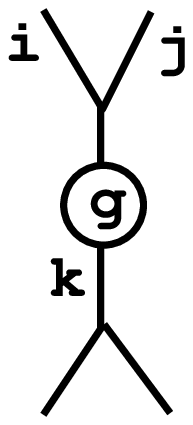}{50}{-25}{2}{2}
\end{equation}
Using this relation one can find the result of the
integral of a product of 3 and 4 group elements.
The corresponding formulas are:
\begin{eqnarray}
\pic{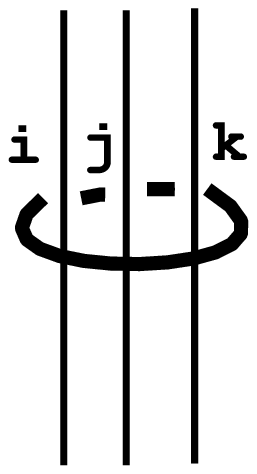}{50}{-25}{2}{2}&=&\pic{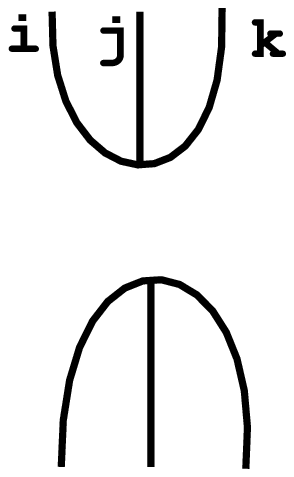}{50}{-25}{2}{2} \\
\pic{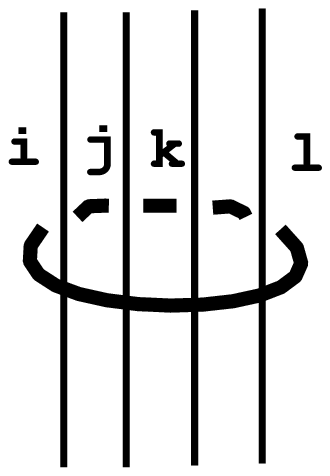}{50}{-25}{2}{2}&=&\sum_m {\rm dim}_m
\pic{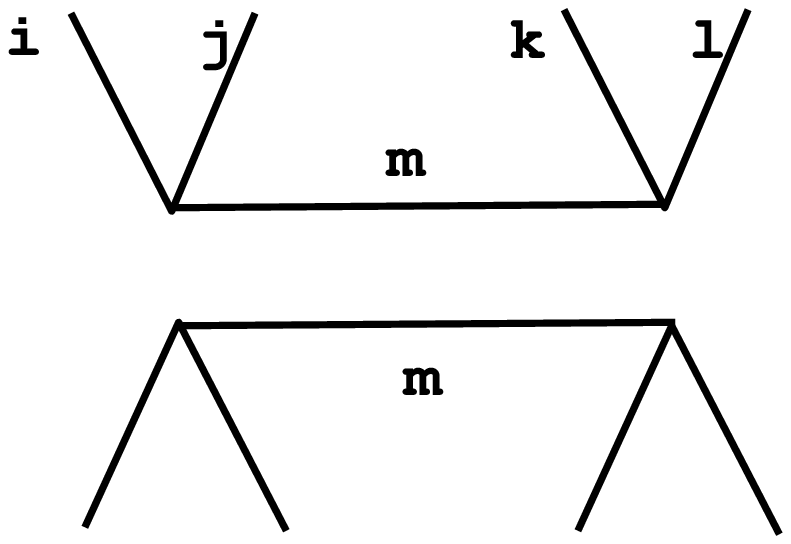}{70}{-35}{2}{2}
\end{eqnarray}

Also, we use the same trivalent vertex to define the normalized
(6j),(15j)-symbols used in the body of the paper. Both symbols
are given by evaluations of the corresponding spin networks,
where the intertwiner used in a trivalent vertex is always
the one normalized as in (\ref{theta-norm}). Thus, we get:
\begin{eqnarray}\label{6j-def}
(6j) = \left(\pic{tet}{60}{-30}{-2}{2}\right)_0 \\ \label{15j-def}
(15j) = \left(\pic{4simpl-nongrasped}{60}{-30}{0}{2}\right)_0,
\end{eqnarray}
where the resolution of the 4-valent vertex in (\ref{15j-def})
is given by:
\begin{equation}
\pic{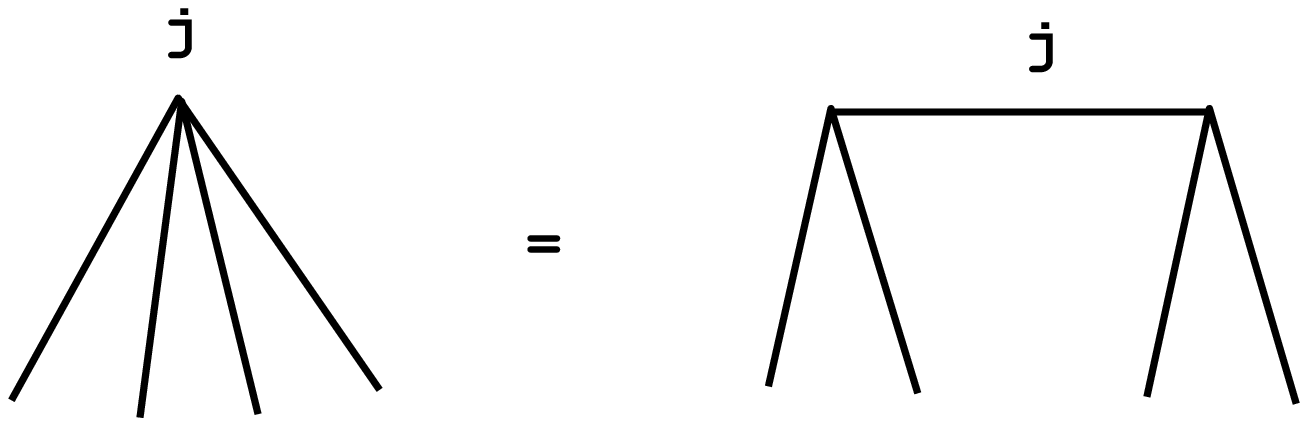}{60}{-30}{2}{2}
\end{equation}

\section{Quantum 6j symbol}

\begin{figure}[htbp]
\centerline{
\psfig{figure=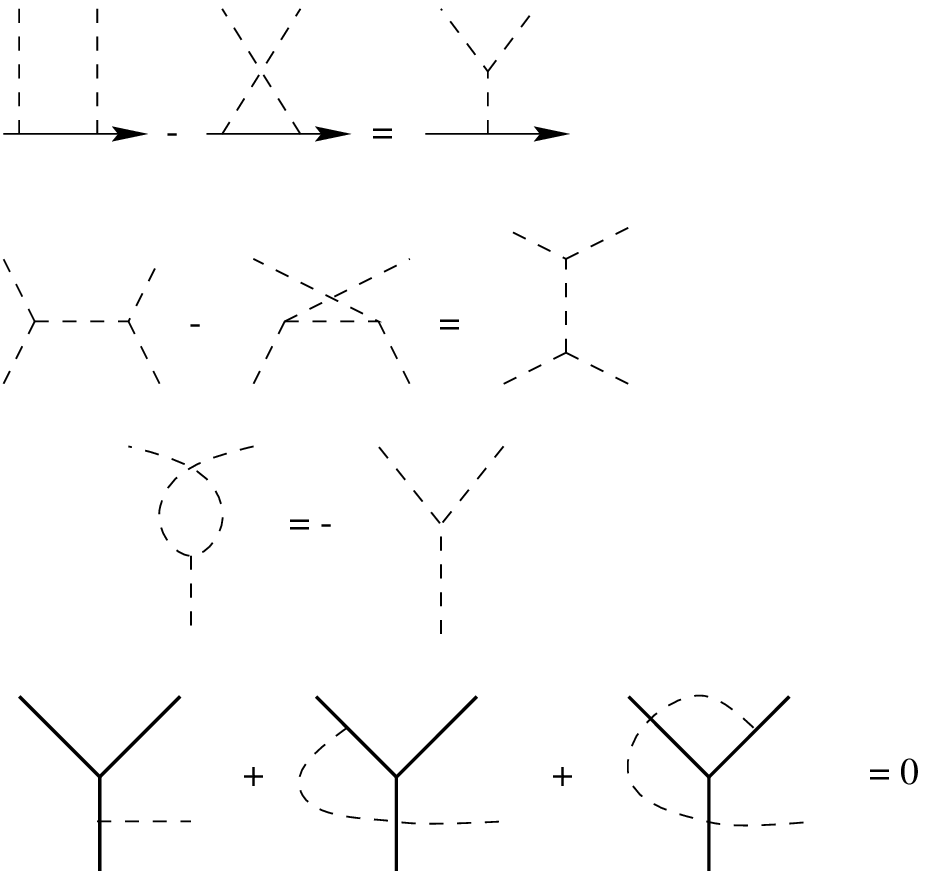,height=3.0in}}
\caption[]{}
\label{ihx}
\end{figure}

Let $X$ be a one dimensional oriented compact manifold, or more generally 
an oriented graph.
A chord diagram (usually referred as Chinese character chord diagram)
with support $X$ 
is the union $D =\bar{D}\cup X$ where $\bar{D}$ is a graph with univalent
and trivalent vertices, together with a cyclic orientation of trivalent
vertices and such that univalent vertices lie in $X$.
Trivalent vertices are referred to as internal vertices,
and the degree of $D$ denoted by 
$d^{\circ}(D)$ is half the number of vertices of
the graph $ \bar{D}$.
Let $\tilde {\cal A}_n$ be the $\bf Z$ module freely generated by 
chord diagrams of degree $n$.
We define the $\bf Z$ module of Vassiliev diagrams of degree n
denoted by ${\cal A}_n$ as being the quotient of $\tilde {\cal A}_n$
by the relations (STU, IHX, AS) shown in Figure \ref{ihx}. In the figures we 
always represent the support $X$ with bold lines and the graph $\bar{D}$ with
 dashed chords.

\begin{figure}[htbp]
\centerline{
\psfig{figure=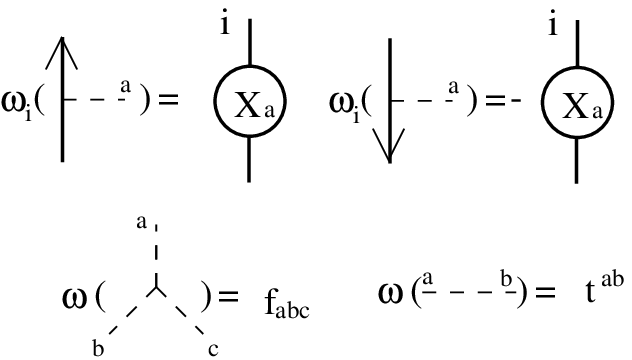,height=2.0in}}
\caption[]{}
\label{eval}
\end{figure}

Given a Lie group $G$, and a coloring $C$ of the graph 
 it is possible to define an evaluation of 
chord diagrams that is usually called weight system and denoted 
by $\omega_{{G,C}}$.
Here $C$, the coloring of the graph $X$, is a map which assigns
 a representation of $G$ to any edge of the graph and an intertwiner to 
any vertex of the graph $X$.
Given a chord diagram  $D$ with support $X$ and a coloring $C$ of $X$,
we define $\omega_{{G,C}}(D)$ as shown in Figure \ref{eval}.
Here $X_a$ denotes a basis of the Lie algebra of $G$, $f^{abc} $ 
denotes the structure constants associated with each internal vertices and 
$t= t^{ab} X_{a}\otimes X_{b}$ is the quadratic casimir $t= t^{ab} 
X_{a}\otimes X_{b}$ associated to each chord.

The space of chord diagrams has been used to define the now famous 
universal invariant of framed link $Z$ \cite{Kon}, which assigns to any 
link $L$ in $R^{3}$ a formal power series in $\hbar$ 
$Z(L)=\sum_{n}\hbar^{n} Z_n(L)$, with $Z_n(L) \in {\cal A}_n$.  We are 
not going into the details of this construction.  However we must say 
that this construction uses three main building blocks: The 
braiding $R\in {\cal A}(I^{2})$, the associator $\phi \in {\cal 
A}(I^{3})$, which should satisfy the pentagon and hexagon identity, 
and the value of the un-knot $\nu \in {\cal A}(I)$, where $I$ denotes 
the unit interval (\ref{expan}) \cite{Dri,Car}.
\begin{eqnarray}
\label{expan}
\phi & = &\kd{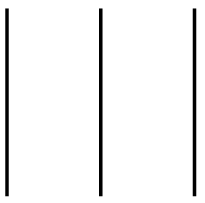} - {\hbar ^2\over 12} \kd{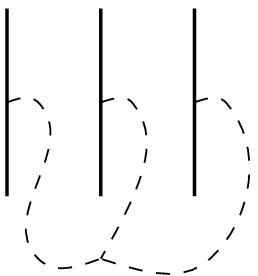}+\cdots 
\\ \label{r-matrix}
R= & =& \kd{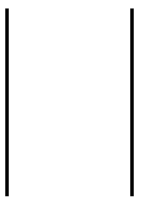} + {i\hbar\over 2} \kd{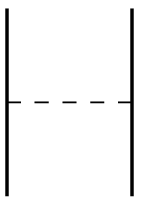} - {\hbar^2 \over 2} 
\kd{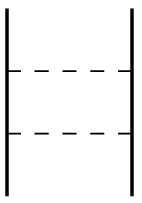}+\cdots 
\\ \label{nu}
\nu &=& \kd{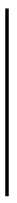} - {\hbar^2 \over 12 } \kd{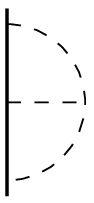} +\cdots
\end{eqnarray}
One important property of the universal invariant is that 
 $\omega_{G,C}(Z(L))$ is equal to the Reshetikhin-Turaev evaluation of 
 the link $L$ associated with the quantum group $U_{q}(G)$ 
 \cite{AF1,Car,LeM1}, where $q =\exp i\hbar$ and the casimir is normalized 
 such that the norm of long roots is $2$.
  
 It has been shown recently \cite{Mur,Fr}, that the universal invariant $Z$ 
  can be extended to an invariant of oriented, framed trivalent graphs 
  embedded into $R^{3}$.  The evaluation of the tetrahedral graph with 
  the universal invariant is given by equation (\ref{tet}).
   \begin{equation}
\label{tet}
Z(\kd{tet})= \KD{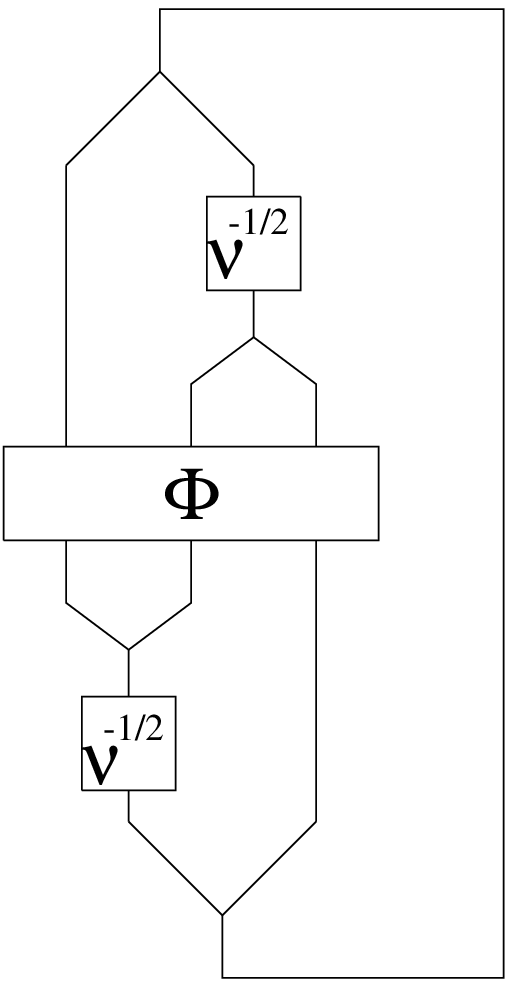}
 \end{equation}
If we evaluate the result of Fig. \ref{tet}
using the weight system $ \omega_{su(2),C}$ we get the normalized 
$6j$ symbol expanded in terms of $\hbar$ (Reshetikhin-Turaev 
evaluation of the tetrahedral graph for $U_{e^{i\hbar}}(su(2))$).
The term proportional to $\hbar^{0}$ is just the classical $6j$
symbol. The next term, proportional to 
$\hbar^{2}$, is given by~:
\begin{equation}
V= {1\over 24}(\kd{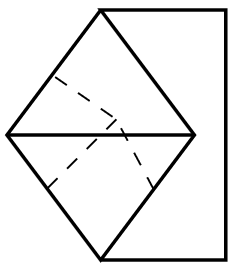} - \kd{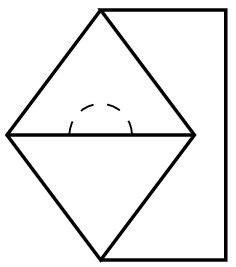} - \kd{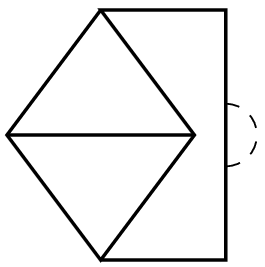}).  \end{equation}
Here we have 
used the fact that, in the case of $su(2)$ and for the normalization of the 
casimir given by the trace in the fundamental representation, the 
identity $\kd{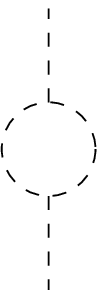}= 4 \kd{grasp2}$ is satisfied.  This term can 
be written in a symmetric form as:
\begin{equation}
\label{tet1}
tet^{(1)}={1\over 16} 
({1\over 24} \sum_{e,e',e"} \langle{\textstyle \KD{grasp}}| 
{\textstyle \KD{tet}}\rangle - {1\over 4} \sum_e \langle
{\textstyle \KD{grasp2}}| \textstyle{\KD{tet}}\rangle),
\end{equation}
where the first sum is over all triples of distinct edges of the tetrahedra
and the second sum is over all edges of the tetrahedra.

\section{Geometric 4-simplex}

In this Appendix we list some facts about the geometry of 
a 4-simplex in $\R^4$. The geometrical considerations used in 
(\ref{sec:bf4d}) are based on some of these facts.

A 4-simplex in $\R^4$ is characterized (up to translations) by
four vectors. These, for instance, can be vectors pointing from
one of the vertices, which we will denote by (0), to the other 
four vertices (1)-(4). See Fig. 10.
\begin{figure}
\centerline{\hbox{\epsfig{figure=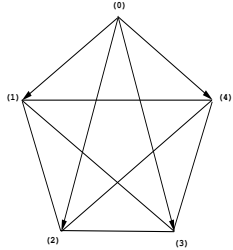,height=1.3in,width=1.2in}}}
\caption{A 4-simplex is characterized up to translations by
four vectors.}
\end{figure}
Let us denote these vectors by $e^I_a$. Here $I=1\ldots 4$
is an index for a vector in $\R^4$, and 
$a=1\ldots 4$ indicates a vertex at which the vector is directed.
Thus, $e^I_1$ is a vector pointing from vertex (0) to the
vertex (1). 

Instead of vectors it is sometimes more convenient to use
the so-called bivectors. A bivector $E^{IJ}$ is an element of the
second exterior power of $\R^4$. In particular, it can be obtained by wedging
two vectors. Thus, bivectors that characterize a 4-simplex
are obtained as $E_{ab}^{IJ} = e^{[I}_a e^{J]}_b$. Here the
brackets denote the operation of antisymmetrization. It is
not hard to see that bivectors $E_{ab}^{IJ}$ are in one-to-one
correspondence with faces of the 4-simplex $h$. For example,
the bivector $E_{12}^{IJ}$ corresponds to the face whose
vertices are (0),(1),(2). The norm of each bivector 
is proportional to the squared area of the corresponding face
\begin{equation}\label{bi-area}
E_{ab}^{IJ} E_{ab\,IJ} = 2 A_{ab}^2,
\end{equation}
where $A_{ab}$ is the area of the face (0),(a),(b) and no summation
over $a,b$ is assumed. The volume of $h$ can be obtained by
wedging two bivectors that correspond to faces that do not
share an edge. For example, with our choice of orientation
of $\cal M$, the volume is given by
\begin{equation}\label{bi-volume}
V_h = {1\over 4!} \epsilon_{IJKL} E_{12}^{IJ} E_{34}^{KL}.
\end{equation}

It is sometimes convenient to introduce bivectors corresponding
to all 10 faces of $h$. So far we have introduced 6 bivectors
$E_{ab}^{IJ}$ corresponding to 6 faces of $h$. 
Bivectors that correspond to other 4 faces can be obtained as 
linear combinations of these 6 bivectors. When working with 
all 10 bivectors, it is convenient to label bivectors by
3 different indices instead of just two. We will employ for 
this purpose small Greek letters. Thus, bivectors are
denoted by $E_{\alpha\beta\gamma}^{IJ}, \alpha,\beta,\gamma=0\ldots 4$.
These bivectors are defined by
\begin{equation}\label{bi}
E_{\alpha\beta\gamma}^{IJ} = e^{[I}_{\alpha\beta} e^{J]}_{\alpha\gamma},
\end{equation}
where $e^I_{\alpha\beta}$ is a vector that points from the
vertex $\alpha$ to vertex $\beta$. The norm of all
bivectors (\ref{bi}) is equal to the twice of the squared
area of the corresponding face, as in (\ref{bi-area}).
As we have said above, only 6 of 10 bivectors (\ref{bi})
are independent. Thus, there are certain relations between
them. One can write one such relation for each tetrahedron
of $h$. One gets 5 relations only 4 of which are independent.
With our definition (\ref{bi}) these relations are
\begin{eqnarray}\label{closure}
E_{012}+E_{023}-E_{013}-E_{123}=0, \\ \nonumber
E_{013}+E_{034}-E_{014}-E_{134}=0, \\ \nonumber
E_{024}+E_{234}-E_{023}-E_{034}=0, \\ \nonumber
E_{014}+E_{124}-E_{012}-E_{024}=0, \\ \nonumber
E_{123}+E_{134}-E_{124}-E_{234}=0, 
\end{eqnarray}
where we have suppressed the indices $I,J$ for brevity.
The volume of $h$ can be expressed as a wedge product of
any two of the bivectors (\ref{bi}) corresponding
to faces that do not share an edge. This can be written as
\begin{equation}\label{bi-volume1}
{\rm sign}(f,f') V_h = {1\over 4!} \epsilon_{IJKL} E(f)^{IJ} E(f')^{KL},
\end{equation}
where we have introduced a notation $E(f)^{IJ}$ for a bivector
that corresponds to face $f$, and ${\rm sign}(f,f')$ is the
sign of the right-hand-side in (\ref{bi-volume1}). The
expression (\ref{bi-volume1}) gives the volume of $h$ for
any two pairs of faces $f,f'$ that do not share an edge.
One can use the 
expression (\ref{bi-volume}) and the ``closure'' relations
(\ref{closure}) to work out the correct sign of any of such 
formula for the volume.

Any bivector can naturally be split in its self-dual and
anti-self-dual parts: 
\begin{equation}
E_{\alpha\beta\gamma}={}^+E_{\alpha\beta\gamma}+{}^-E_{\alpha\beta\gamma}.
\end{equation}
The self-dual and anti-self-dual parts are given correspondingly by
\begin{eqnarray}
{}^+E_{\alpha\beta\gamma}={1\over2}(E_{\alpha\beta\gamma}+
{}^*E_{\alpha\beta\gamma}), \\ \nonumber
{}^-E_{\alpha\beta\gamma}={1\over2}(E_{\alpha\beta\gamma}-
{}^*E_{\alpha\beta\gamma}),
\end{eqnarray}
where the Hodge star duality operation is defined as
\begin{equation}
{}^*E_{\alpha\beta\gamma}^{IJ} = {1\over2} \epsilon^{IJ}_{KL}
E_{\alpha\beta\gamma}^{KL}.
\end{equation}
Since the space of self-dual (and anti-self-dual) bivectors
in $\R^4$ is three-dimensional, we can introduce a new 
set of indices to label them. Thus, as the index for
self-dual (anti-self-dual) bivector we will use lower case Latin
letters from the middle of the alphabet: $i,j,k,\ldots = 1,2,3$.
The norm of any self-dual (anti-self-dual) bivector calculated
by contracting indices $I,J$ will be the same as the norm
calculated by contracting the single index $i$:
\begin{equation}
{}^+E^i {}^+E_i = {}^+E^{IJ} {}^+E_{IJ},
\end{equation}
where we have suppressed the indices $\alpha,\beta,\ldots$.

Not any bivector in $\R^4$ is simple, that is, not any bivector
is a wedge product of two vectors. The necessary and sufficient
requirement of simplicity is that the norm of the self-dual
part is equal to the norm of the anti-self-dual part of the 
bivector. Thus, using (\ref{bi-area}), we
can conclude that when a bivector is simple, its 
self-dual part norm is equal to the squared area
of the corresponding face.
\begin{equation}\label{biselfarea}
{}^+E_{\alpha\beta\gamma}^i {}^+E_{\alpha\beta\gamma\,i} = 
A_{\alpha\beta\gamma}^2.
\end{equation}
There exists an expression for the volume of $h$
that involves only the self-dual parts of bivectors (\ref{bi}):
\begin{equation}\label{sd4volume}
V_h = {1\over 3!} {1\over 30}\sum_{f,f'} {\rm sign}(f,f') {}^+E^i(f) {}^+E_i(f'),
\end{equation}
where ${}^+E^i(f)$ is the self-dual part of the bivector
(\ref{bi}) corresponding to the face $f$, the sum is taken over
all pairs $f,f'$ of faces that do not share an edge, and 
${\rm sign}(f,f')$ is the function introduced above by
equation (\ref{bi-volume1}). The factor of $1/30$ in (\ref{sd4volume})
appears to cancel the factor that comes from the sum
over 30 terms $f,f'$.

\end{document}